\documentclass[a4paper,11pt]{article}
\usepackage{pos}
\usepackage{graphicx}
\usepackage{mathtools}
\usepackage{color}
\definecolor{blue}{rgb}{0.1,0.2,0.8}
\definecolor{purple}{rgb}{0.6,0,0.5}
\usepackage{epsfig}
\usepackage{longtable}
\usepackage{nicefrac}
\usepackage{bm}
\usepackage{tabularx}

\newcolumntype{L}[1]{>{\raggedright\arraybackslash}p{#1}}
\newcolumntype{C}[1]{>{\centering\arraybackslash}p{#1}}
\newcolumntype{R}[1]{>{\raggedleft\arraybackslash}p{#1}}

\renewcommand{\vv }{\mathrm{v}}

\title{Heavy meson chiral Lagrangians, effective flavored couplings, SU(4) flavor breaking and their consequences}
 \ShortTitle{Heavy meson chiral Lagrangians, effective couplings and SU(4) flavor breaking }

\author*[a]{Bruno El-Bennich}
\author[a,b]{Fernando E. Serna}

\affiliation[a]{Laborat\'orio de F\'isica Te\'orica e Computacional, Universidade Cidade de S\~ao Paulo,\\
                    Rua Galv\~ao Bueno 868, 01506-000 S\~ao Paulo, SP, Brazil}

\affiliation[b]{Departamento de F\'isica, Universidad de Sucre,  \\ 
                      Carrera 28 No. 5-267, Barrio Puerta Roja, Sincelejo, Colombia}

\emailAdd{bruno.bennich@cruzeirodosul.edu.br}
\emailAdd{fernando.enrique@unesp.br}

\abstract{We review heavy quark flavor and spin symmetries, their exploitation in heavy meson effective theories and the flavored couplings of charmed and light mesons
               in the definition of their effective Lagrangians. We point out how nonperturbative continuum QCD approaches based on Dyson-Schwinger and Bethe-Salpeter equations
               can be used to calculate strong and leptonic decays of open-charm mesons and heavy quarkonia. The strong decay $D^*\to D\pi$ serves as a benchmark, as it is the only 
               physical open-charm observable that can be related to the effective Lagrangian's couplings. Nonetheless, a quantitative comparison of $D^*D\pi$, $\rho DD$, 
               $\rho D^*D$ and $\rho D^* D^*$ couplings for a range of off-shell momenta of the $\rho$-meson invalidates SU(4)$_F$ symmetry relations between these couplings.
               Thus, besides the breaking of flavor symmetry by mass terms in the Lagrangians, the flavor-symmetry breaching in couplings and their dependence on the $\rho$-meson
               virtuality cannot be ignored. We also take the opportunity to present new results for the effective $J/\psi DD$ and $J/\psi D^*D$ couplings. We conclude this contribution with 
               a discussion on how the description of pseudoscalar and vector $D$, $D_s$,  $B$ and $B_s$ meson properties can be drastically improved with a modest modification of
               the flavor-dependence in the Bethe-Salpeter equation.      }

\FullConference{%
   10th International Workshop on Charm Physics (CHARM2020),  \\
   31 May - 4 June, 2021 \\
   Mexico City, Mexico -- Online 
}
 
 \tableofcontents

\begin{document}
\maketitle


\section{Introduction}

Heavy flavor physics has been an important and very active field of particle physics for most of the past three decades and is posed to remain a formidable source of new discoveries. Most efforts 
have been dedicated to the study of weak $B$-meson decays and to establishing experimental evidence for direct and/or indirect (in oscillation) CP-violation. In the 2000s, the BaBar experiment 
at the Stanford Linear Accelerator Center (SLAC) and the Belle Experiment at the High Energy Accelerator Research Organization (KEK), both electron-positron colliders with the center of mass energy 
tuned to the $\Upsilon (4S)$ resonance, were very successful in observing a large number of CP-violating processes in $B$ decays. In the following decade dedicated experiments were pursued by the 
LHC$b$ experiment with the Large Hadron Collider (LHC) at CERN. The new generations of detectors at the LHC provided hitherto unprecedented luminosities, which led to the discovery of CP violation 
in $B_s$ decays~\cite{LHCb:2013syl} and most recently CP violation was reported in $D^0 \to K^-K^+$ and $D^0 \to \pi^- \pi^+$ decays~\cite{LHCb:2019hro}.

These important experimental efforts by BaBar and Belle, and nowadays by the LHC$b$ and Belle II collaborations, have strongly improved our understanding of CP-violating mechanisms. Much
of our current knowledge about  the electroweak sector of the Standard Model will serve as a guideline to interpret the weak interactions of any new particles to be discovered. On the other hand, 
the contributions of the strong force to heavy meson decay amplitudes are still the major source of uncertainty in this field. That is to say, theoretical approaches to weak decays based on perturbative 
Quantum Chromodynamics (QCD), such as heavy quark effective field theory and associated factorization theorems, provide the means to systematically  integrate out energy scales in the perturbative 
domain. This yields approximations of the decay amplitudes in terms of products of hard and soft matrix elements valid in the heavy-quark limit. However, a reliable evaluation of the soft physics, 
namely hadronic wave functions and form factors, is notoriously difficult. These difficulties must be overcome, as the precise knowledge of hadronic contributions to heavy meson decay 
amplitudes is crucial to determine strong phases without which CP violation cannot occur~\cite{El-Bennich:2009gbu,El-Bennich:2006rcn,El-Bennich:2009gqk}. 

Flavor physics is not merely a fertile playground for {\em indirect\/} precision tests of the Standard Model, as it also provides a powerful tool to study nonperturbative aspects of QCD in charm 
and bottom mesons and in heavy quarkonia. Charm physics, on its own right, will be the object of much experimental activity at the Facility for Antiproton and Ion Research (FAIR), at the 
Japan Proton  Accelerator Research Complex (J-PARC) and at the Beijing Spectrometer (BES III). Notably, in this context, the masses and quantum numbers of recently discovered 
$D$ mesons and their possible radial excitations are not yet precisely determined (the preferred interpretations are scalar and vector $D$ and $D_s$ mesons in some cases). The discovery of new
types of charmonia, especially the plethora of observed $X,Y,Z$ states, is exciting as these objects challenge the commonplace view of hadrons as either $\bar qq$ or $qqq$ color-singlet states. 
Their composition remains controversial and numerous theoretical descriptions have been proposed. In case of the $X(3872)$ this ranges from mixtures of pure charmonia with molecular states to
purely molecular bound states and tetraquarks. Molecular states, which are bound states of mesons, are plausible since the mass difference between the $X(3872)$ and the $\bar D^0 D^{*0}$  
threshold is tiny. 

The effective Lagrangian approaches we discuss in Section~\ref{sec2} treat the observed decays, for instance $X(3872)\to J/\psi\,\pi^+\pi^-$~\cite{Belle:2003nnu, BaBar:2004oro,CDF:2003cab,
D0:2004zmu}, in terms of mesonic degrees of freedom, where a $J/\psi\, \rho$ state in $X(3872)\to J/\psi\, \rho \to J/\psi\, \pi^+\pi^-$ is preceded by a $D$-meson loop that couples to the $J/\psi$ 
and  $\rho$~\cite{Liu:2006df}.  Similarly, the reaction $D^{(*)}D^{(*)} \to \pi X(3872)$ is effectively described by an intermediate triangle diagram with pseudoscalar and vector 
mesons~\cite{MartinezTorres:2014son}.  As we will see, the effective Lagrangians are expressed in terms of effective flavored couplings between $D$- and light mesons  which are not
known \emph{a priori}.

\begin{figure}[t!]
\centering
  \includegraphics[scale=0.6]{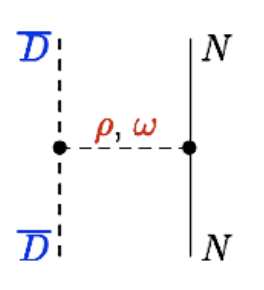} \hfill 
  \includegraphics[scale=0.6]{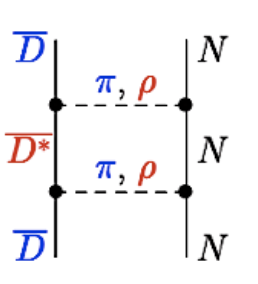} \hfill
  \includegraphics[scale=0.6]{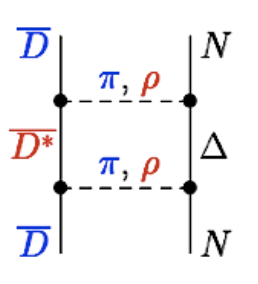} \hfill
  \includegraphics[scale=0.6]{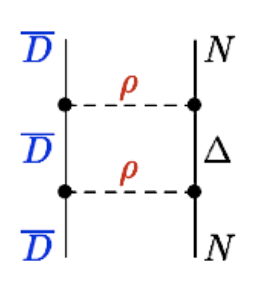}  
  \caption{A set of representative yet not exhaustive diagrams of  $DN$ scattering via the exchange of light pseudoscalar and vector mesons. The $\pi NN$ coupling can be
  be deduced from the experimental pion-nucleon scattering lengths of pionic atoms~\cite{Ericson:2000md} and the $D^*D\pi$ coupling can be extracted from the physical decay width of $D^{*+} \to D^0\pi^+$. 
  It is, however, misguided to assume SU(4)$_F$ flavor symmetry expressed by relations of the kind $g_{ D^* D\pi}  = g_{\rho D D}=g_{\rho D^* D^*}$ in applications of effective Lagrangians to these 
  scattering amplitudes. At which light-meson virtuality $q^2$ the couplings are evaluated should also be taken into consideration.} 
\label{DNdiagrams}
\end{figure}

Related considerations apply to the prospects of studying in-medium $D$ mesons, which may open the possibility of forming novel charmed nuclear bound states. Their existence is contentious, 
yet the study of interactions between charmed mesons and nuclear matter represents an important component of the proposed $\overline{\textrm P}$ANDA activities at FAIR, where low-momentum 
charmonia, such as $J/\psi$ and $\psi$, as well as $D^{(\ast )}$ mesons, will be produced by antiproton annihilation on nuclei~\cite{Haidenbauer:2014rva}. Effective Lagrangians have also been 
employed to calculate $DN$ cross section with the diagrams depicted in Figure~\ref{DNdiagrams}. They are typically formulated with couplings between $D^{(\ast)}$ mesons and light pseudoscalar 
and vector mesons and are derived from an SU(4)$_F$  extension of light-flavor chirally-symmetric Lagrangians. 

At a future Electron-Ion Collider (EIC),  the electroproduction of these charmonia from nuclei at threshold may be exploited to reveal the existence of  hidden-charm pentaquarks~\cite{LHCb:2015yax}  
and offers prospects for a quantitative estimate of the QCD trace anomaly~\cite{Kharzeev:1995ij,Meziani:2020oks,Krein:2020yor}. Irrespective of the experiment's aim, the extraction of valuable 
information from cross sections implies the difficult task of understanding charmonium production and its final-state wave function~\cite{Xu:2021mju} whose light-front projections were recently 
computed in Ref.~\cite{Serna:2020txe} and are discussed in another contribution to these proceedings~\cite{Serna:2021xnr}. 

Thus, while we just swept a wide range of physical motivations and experimental and theoretical challenges in heavy flavor physics, their strong appeal to a hadron physicist should be evident. 
In the following, we will review various aspects of charm physics and the motivation for heavy flavor symmetry in effective Lagrangians, while also pointing out a possible misguidance in assuming 
universal  couplings between heavy and light mesons. Flavor-symmetry breaking and a nonperturbative description of mesons are indeed the common thread of this contribution.


\section{Flavor and spin symmetries in effective theories  \label{sec2}  }

In an effective heavy-quark expansion of QCD it is common practice to take the limit $m_Q\to \infty$ as a good approximation for quarks with masses $m_Q \gg \Lambda_\mathrm{QCD}$.
This leads to heavy-spin and -flavor symmetries which are \emph{not\/} manifest in the QCD Lagrangian but emerge in heavy quark effective theory (HQET)~\cite{Neubert:1993mb,Manohar:2000dt,
El-Bennich:2012hom}.  In a nutshell, this effective theory consists in an expansion in powers of $m_Q^{-1}$ and $\alpha_s$ and to leading order the HQET Lagrangian reads,
\begin{equation}
  \mathcal{L}_\mathrm{HQET} \ = \  \bar h_\vv \, i \vv \cdot D\, h_\vv \  + \ \frac{1}{2m_Q}  \Big [ \bar h_\vv  \left (i\,{\vec D} \right )^{\!2} \! h_\vv 
      + c(\mu) \dfrac{g}{2}\, \bar h_\vv \, \sigma_{\mu\nu}\, t^a G_{\mu\nu}^a \, h_\vv   \Big ]  \ + \ \   \mathcal{O}\left ( m_Q^{-2}, \alpha_s^2 \right ) \   ,
 \label{lagrangeHQET}
 \end{equation}
where $D_\mu = \partial_{\mu}+g t^a A_{\mu}^{a}$ is the usual covariant derivative, $A^a_\mu$ is the gluon field, $G_{\mu\nu}^a$ is the gluon field strength tensor and $t^a = \lambda^a/2$  are the 
SU(3) color group generators. The Wilson coefficient $c(\mu)$ represents higher-order operator corrections in $\alpha_s(\mu) =g^2/4\pi$ with a matching  condition at tree level that implies $c(m_Q) =1 
+ \mathcal{O}[\alpha_s(m_Q^2)]$ at the renormalization point $\mu=m_Q$.  The heavy quark interacts with the light quark by exchanging momenta $k$ much smaller than its  mass, i.e. $k \sim 
\Lambda_\mathrm{QCD} \ll m_Q$. Hence, being almost on-shell, it  moves with the hadron velocity\footnote{\, The four-velocity satisfies $\vv^2=1$ with $\vv_\mu = (1,\vec 0)$ 
in the rest frame of the heavy quark.} $\vv_\mu$ and to a good approximation the heavy-quark momentum can be written as, 
\begin{equation}
   p_\mu = m_Q \vv_\mu +k_\mu \ , 
   \label{heavymomentum}
\end{equation}
close to the momentum $m_Q \vv_\mu$ of the hadron's motion. In other words, in the limit $m_Q\to \infty$, the heavy-quark velocity becomes a conserved quantity and the momentum exchange with 
surrounding light constituents is predominantly soft. In the Lagrangian~\eqref{lagrangeHQET}, $h_\vv$ denotes the large component of the Dirac spinor field obtained from the positive-energy projection, 
\begin{equation}
 h_\vv (x) = e^{im_Q \vv \cdot x}\ \frac{1+ \gamma\cdot \vv }{2}\ Q(x)\ , 
\end{equation}
where $Q(x)$ is the heavy-quark spinor in the QCD Lagrangian,  $\mathcal{L}_Q = \bar Q (\gamma \cdot D - m_Q) Q $. The exponential phase factor serves to subtract $m_Q \vv_\mu$ from the heavy 
quark momentum, so that in momentum space a derivative acting on $h_\vv$(x) produces the residual momentum $k$. The small component of the quark spinors  is suppressed in $m_Q^{-1}$ and integrated 
out in the effective theory. 

In the infinite-heavy quark limit only the leading term survives in the Lagrangian~\eqref{lagrangeHQET}, which exhibits the full $U(2N_h)$ spin-flavor symmetry ($N_h$: heavy flavor number): it is 
mass-independent and invariant under rotations in flavor space, and in the absence of Dirac matrices heavy-quark interactions with gluons leave its spin unchanged while $h_\vv(x)$ is invariant 
under spin rotation. When applying Feynman rules it gives rise to the well known quark propagator in the heavy-quark limit, which can also be derived using Eq.~\eqref{heavymomentum}: 
\begin{equation}
   S(p) \, = \, i \, \frac{\gamma\cdot p + m_Q}{p^2 - m_Q^2}\ \  \xrightarrow{m_Q\to \infty} \ \ i\, \frac{1+\gamma\cdot \vv }{2\, \vv \cdot k}\ + \ \mathcal{O}  \left (\frac{k}{m_Q} \right ) \ .
 \label{mqpropagator}
\end{equation}

Since the heavy-quark spin decouples in this limit, the light quarks do not experience any different interactions with a heavy constituent quark in a pseudoscalar or in vector meson. 
This implies symmetry relations between the decay constants of heavy mesons: $f_D=f_{D^*}$  and $f_B=f_{B^*}$. Of course, in full QCD this is not even the case for the $B$-mesons. 
The second term in Eq.~\eqref{lagrangeHQET} is the kinetic energy, ${\vec p\,}^2\!/2m_Q$, of a heavy non-relativistic constituent quark  and breaks the heavy quark flavor symmetry.
Both, heavy quark spin and flavor symmetries are broken by $\mathcal{O}(\Lambda_\mathrm{QCD}/m_Q)$ and $\alpha_s$ corrections due the chromomagnetic term in Eq.~\eqref{lagrangeHQET}, 
and so do higher terms in the  expansion as well as nonperturbative contributions. More precisely, gluon exchanges renormalize coefficients that multiply effective operators.
For example, the renormalization scale dependence of $c(\mu)$ must be cancelled by that of the chromomagnetic moment operator. However, in evaluating matrix elements with initial 
and final hadron states, the operators give rise to form factors which break flavor symmetry and whose scale dependence is often unknown due to the model approach employed. 
To summarize, HQET can schematically be expressed as the expansion, 
\begin{equation}
  \mathcal{L}_\mathrm{HQET}  \ = \  \mathcal{L}_{0} + \frac{1}{m_{Q}}\, \mathcal{L}_{1}+\frac{1}{m_{Q}^{2}} \, \mathcal{L}_{2} + \, \ldots \ ,
\end{equation}
in which $\mathcal{L}_{0}$ is manifestly spin and flavor independent, whereas the $1/m_Q$ terms are symmetry breaking corrections. 

Spin and flavor symmetries have  also guided the construction of an effective Lagrangian that describes the interactions of the pseudoscalar and vector $D^{(*)}$ and $B^{(*)}$ mesons with 
light pseudoscalar mesons~\cite{Burdman:1992gh,Wise:1992hn,Casalbuoni:1996pg}. This  Lagrangian satisfies C, P, T and Lorentz invariance and at leading order in the $1/M_H$ expansion, 
where $M_H$  is the heavy-meson mass, it  imposes flavor and spin symmetry in the heavy meson sector and chiral $\mathrm{SU} (3)_L \otimes \mathrm{SU} (3)_R $ invariance 
in the light sector:
\begin{equation}
   \mathcal{L}_{\mathrm{HMChPT}} \ =\  - \operatorname{Tr} \big [ \bar{H}_a \, i \vv \cdot D_{ab} H_b \big ] 
                                                               +  i\hat{g}  \operatorname{Tr} \big [ \bar{H}_a  \gamma_{\mu}\gamma_{5}  \mathbf{A}^{ab}_{\mu} H_b \big ] 
                                                               +   \frac{f^2_\pi}{8} \, \partial_{\mu} \xi^2_{a b} \partial_{\mu} \xi_{b a}^{2\dagger}  \ ,             
 \label{HMChPT}                                                                                            
\end{equation}
where the trace is over Dirac and flavor indices, the latter represented by a subscript that denotes the light antiquark in the heavy meson: $H_a =Q\bar q_a$, $a = u,d,s$. Here, $\vv$ are the
heavy-meson velocities and a sum over them is implicit. The covariant derivative is defined as the sum of the partial derivative and the vector current of the light-meson octet, 
\begin{equation}
  D^{ab}_{\mu} H_{b}=\partial_{\mu} H_{a}- \, \tfrac{1}{2}\left[\xi^{\dagger} \partial_{\mu} \xi+\xi \partial_{\mu} \xi^{\dagger}\right]_{ab} H_{b} \ ,
\end{equation}
and has the right transformation properties under chiral transformations
with the  negative parity spin doublets, $D, D^*$ and $B, B^*$, represented by Dirac tensor structures as,
\begin{equation}
 H_a (\vv) =  \frac{1+\gamma\cdot \vv}{2}  \left [  P_{\mu}^{* a}(v) \gamma_{\mu}-P^{a}(v) \gamma_{5} \right ]  \ , \quad  \bar H_a =  \gamma_0 H_a^\dagger \gamma_0 \ .
\end{equation}
The heavy vector and pseudoscalar fields are described by annihilation operators, $P^a$ and $P^{*a}_\mu$, which are normalized as:
\begin{align}
    \langle 0 | P^a | H_a (0^{-} ) \rangle &= \sqrt{M_H}  \ , \\
    \langle 0\left |  P^{*a}_\mu \right | H^*_a  (1^{-}) \rangle & = \sqrt{M_H} \, \bm{\epsilon}^\lambda_{\mu}  \ .
\end{align}
One has by definition $\vv\cdot P^{*a} = 0$, $ \bm{\epsilon}^\lambda$ is the polarization of the vector meson and $M_P = M_{P^*} = M_H$ at this leading order. 
The light-meson octet also enters the Lagrangian via the axial current $\mathbf{A}_{\mu}^{a b}$,
\begin{equation}
     \mathbf{A}_{\mu}^{ba}=\frac{i}{2}\left[\xi^{\dagger} \partial_{\mu} \xi-\xi \partial_{\mu} \xi^{\dagger}\right]_{ba} \ ,
\end{equation}
which transforms as the adjoint representation of SU(3)$_V$ and makes use of the exponential representation $\xi=\exp \left(i \mathcal{M} / f_{\pi} \right)$, where $f_\pi$ is the weak decay 
constant of the pion in the chiral limit and $\mathcal{M}$ is the $3 \times 3$  hermitian, traceless matrix of the pseudo-Goldstone boson octet:
\begin{equation}
  \mathcal{M}  =  \left ( 
     \begin{array}{ccc}
           \sqrt{\frac{1}{2}}\, \pi^{0}+\sqrt{\frac{1}{6}}\, \eta & \pi^{+} & K^{+}   \\
          \pi^{-} & -\sqrt{\frac{1}{2}}\, \pi^{0}+\sqrt{\frac{1}{6}}\, \eta & K^{0}   \\
           K^{-} & \bar{K}^{0} & - \sqrt{\frac{2}{3}}\,  \eta
      \end{array}  \right  ) \ .
\end{equation}
Their coupling to the heavy mesons is determined by a putative \emph{ universal coupling constant\/} $\hat g$. The last term in Eq.~\eqref{HMChPT} is the non-linear Lagrangian that describes
the light meson self-interactions. The interactions between heavy and light mesons are obtained by expanding the field $\xi$ and taking traces. Corrections to the Lagrangian stem from 
higher terms in the $1/M_H$ expansion and from chiral symmetry breaking; for details we refer to  Ref.~\cite{Casalbuoni:1996pg}.

While SU$(3)_F$ is a sensible approximation in hadron physics, though broken at the $\approx 20\%$ level, it is questionable whether this is still reasonable for SU$(4)_F$. At the quark level,  
we saw that  at next-to-leading order in the effective Lagrangian~\eqref{lagrangeHQET}  flavor and spin symmetries are violated and similarly so for charmed mesons in the Lagrangian of Heavy 
Meson Chiral Perturbation Theory (HMChPT) of Eq.~\eqref{HMChPT}. However, the charm and $D$ masses, unlike those of the $b$-quark and $B$-mesons, sit in an uncomfortable energy region.
They are not light, nor are they heavy enough to neglect higher-order $1/M_H$ corrections. Moreover, it is assumed that the coupling $\hat g$ between the light and heavy meson fields is universal 
at leading order in the Lagrangian~\eqref{HMChPT}, i.e. it can be obtained from the strong $B^*B\pi$ coupling with renormalization of $\hat g$ by $\mathcal{O}(1/M_H)$ corrections~\cite{Casalbuoni:1996pg}. 
However, as we will see in Section~\ref{sec3}, it makes a material difference whether $\hat g$ is extracted from a hadronic $g_{D^*D\pi}$ or  $g_{B^*B\pi}$ coupling. The former coupling can be related 
to the physical decay $D^*\to D\pi$~\cite{CLEO:2001sxb,BaBar:2013zgp}, though the latter is unphysical considering the mass difference $m_{B^*} - m_B < m_\pi$. Still, one can calculate 
$g_{B^*B\pi}$  in the chiral limit,  $m_\pi \to 0$.  

Besides HMChPT, other like-minded approaches to interactions between heavy and light mesons are based on a phenomenological ``bottom-up'' SU$(4)_F$ Lagrangian, not derived from a $1/m_Q$ 
expansion of the heavy-quark spinor and tantamount to a meson-exchange model. In the limit of SU$(4)_F$ invariance,  the free Lagrangian for pseudoscalar and vector 
mesons is given by~\cite{Matinyan:1998cb,Lin:1999ad,Oh:2000qr,Bracco:2011pg},
\begin{equation}
  \mathcal{L}_{0} \, =\,\operatorname{Tr}\big (\partial_{\mu} P^{\dagger} \partial_{\mu} P\big )-\frac{1}{2} \operatorname{Tr}\left(F_{\mu \nu}^{\dagger} F_{\mu \nu}\right) \ ,
  \label{L0}
\end{equation}
where $F_{\mu\nu} = \partial_\mu V_\nu - \partial_\nu V_\mu$, and $P$ and $V$ denote the $4 \times 4$ pseudoscalar and vector meson matrices in SU$(4)_F$ with proper normalization:
\begin{equation}
 P  =  \frac{1}{\sqrt{2}} \left( 
 \begin{array}{cccc}
    \frac{\pi^{0}}{\sqrt{2}}+\frac{\eta}{\sqrt{6}}+\frac{\eta_{c}}{\sqrt{12}} & \pi^{+} & K^{+} & \bar{D}^{0} \\
    \pi^{-} & -\frac{\pi^{0}}{\sqrt{2}}+\frac{\eta}{\sqrt{6}}+\frac{\eta_{c}}{\sqrt{12}} & K^{0} & D^{-} \\
    K^{-} & \bar{K}^{0} & -\sqrt{\frac{2}{3}} \eta+\frac{\eta_{c}}{\sqrt{12}} & D_{s}^{-} \\
     D^{0} & D^{+} & D_{s}^{+} & -\frac{3 \eta_{c}}{\sqrt{12}}
\end{array} \right ) \ ,
\label{Pmatrix}
\end{equation}
\begin{equation}
 V =  \frac{1}{\sqrt{2}} \left ( 
 \begin{array}{cccc}
   \frac{\rho^{0}}{\sqrt{2}}+\frac{\omega^{\prime}}{\sqrt{6}}+\frac{J / \psi}{\sqrt{12}} & \rho^{+} & K^{*+} & \bar D^{* 0} \\
   \rho^{-} & -\frac{\rho^{0}}{\sqrt{2}}+\frac{\omega^{\prime}}{\sqrt{6}}+\frac{J / \psi}{\sqrt{12}} & K^{* 0} & D^{*-} \\
   K^{*-} & \bar K^{* 0} & -\sqrt{\frac{2}{3}} \omega^{\prime}+\frac{J / \psi}{\sqrt{12}} & D_{s}^{*-} \\
   D^{* 0} & D^{*+} & D_{s}^{*+} & -\frac{3 J / \psi}{\sqrt{12}}  
 \end{array}  \right) \ .
\label{Vmatrix}
\end{equation}
Next, to introduce interactions between pseudoscalar and vector mesons one applies the following minimal prescription:
 \begin{align}
        \partial_{\mu} P  & \ \rightarrow \ \partial_{\mu} P-\frac{i\hat g}{2} \left [ V_{\mu}, P \right ]  \ , \\
        F_{\mu \nu}        & \ \rightarrow \ \partial_{\mu} V_{\nu}-\partial_{\nu} V_{\mu}-\frac{i\hat g}{2} \left [ V_{\mu}, V_{\nu} \right ] \ .
\end{align}
Applying these substitutions in the Lagrangian~\eqref{L0} and taking into account the hermiticity of the matrices, namely $P = P^\dagger$ and $V = V^\dagger$, one arrives at the Lagrangian:
\begin{align}
   \mathcal{L}  \ =  & \ \  \mathcal{L}_{0} + i\hat g \operatorname{Tr}  \left ( \partial_{\mu} P\left[P, V_{\mu}\right]\right )-\frac{\hat g^{2}}{4} \operatorname{Tr}\left(\left[P, V_{\mu}\right]^{2} \right ) \nonumber \\
                         & + \   i\hat g \operatorname{Tr} \left ( \partial_{\mu} V_{\nu}\left[V_{\mu}, V_{\nu}\right]\right ) + \frac{\hat g^{2}}{8} \operatorname{Tr}\left(\left[V_{\mu}, V_{\nu}\right]^{2} \right ) \ .
 \label{phenLagrangian}                         
\end{align}
In this approach, SU$(4)_F$ flavor symmetry is explicitly broken by hadron-mass terms we here omitted, as they are irrelevant to the discussion, and for which experimental
values are used.  However, due to the flavor symmetry imposed, some exact relations between the couplings exist as a consequence of group properties, as will be seen shortly. 
The Lagrangian in Eq.~\eqref{phenLagrangian} accounts for the three-point functions $PPV$ and $VVV$. Yet, to include anomalous parity terms that give rise to $PVV$ vertices, it is customary 
to resort to the gauged Wess-Zumino action and derive an anomalous three-meson Lagrangian~\cite{Oh:2000qr,Braghin:2021qmu}:
\begin{equation}
  \mathcal{L}_\mathrm{an.} =\, -\frac{\hat g_{a}^{2} N_{c}}{16 \pi^{2} f_{\pi}} \, \epsilon_{\mu \nu \alpha \beta} \operatorname{Tr} \left (\partial_{\mu} V_{\nu} \partial_{\alpha} V_{\beta} P \right ) \ .
\label{anomLagrange}  
\end{equation}

For practical applications to specific mesons, the Lagrangians~\eqref{phenLagrangian} and \eqref{anomLagrange} are expanded in terms of the pseudoscalar- and vector-meson matrices 
in Eqs.~\eqref{Pmatrix} and \eqref{Vmatrix}, where only the entries for the mesons of interest are kept and all other matrix elements are set to zero; see Ref.~\cite{Bracco:2011pg} for further 
details. Specifying to the case of the $\pi$,  $\rho$, $D$ and $J/\psi$ mesons, which is relevant to a phenomenological description of $J/\psi$ suppression in final-state interactions of heavy-ion 
collisions, the relevant interaction Lagrangians derived from Eq.~\eqref{phenLagrangian} and  \eqref{anomLagrange} are:
\begin{spreadlines}{0.7em}
\begin{align}
 \mathcal{L}_{D^*\! D\pi  }    & = \ i g_{ D^{*} Dpi} \,D^*_\mu\, \vec{\tau} \cdot\left(\bar{D} \partial_{\mu} \vec{\pi}-\partial_{\mu} \bar{D} \vec{\pi}\right)  \ , \\
  \mathcal{L}_{D^{*}\! D^{*}\pi}  &  = - g_{D^{*} D^{*}\pi } \, \epsilon_{\mu \nu \alpha \beta} \, \partial_{\mu} D_{\nu}^{*}\vec{\tau} \cdot \vec \pi\, \partial_{\alpha} \bar{D}_{\beta}^ *  \ , \\       
 \mathcal{L}_{\psi D D}     & = \ i g_{\psi D D} \, \psi_\mu \left ( \partial_{\mu} D \bar{D} - D \partial_{\mu} \bar{D}  \right)  \ , \\
 \mathcal{L}_{\psi D^*\! D}  & = \ g_{\psi D^{*} D} \, \epsilon_{\mu \nu \alpha \beta}\, \partial_{\mu} \psi_{\nu}\left(\partial_{\alpha} D_{\beta}^{*} \bar{D}+D \partial_{\alpha} \bar{D}_{\beta}^{*} \right ) \ ,\\
 \mathcal{L}_{\psi D^{*}\! D^{*}} & = \  i g_{\psi D^{*} D^{*}}\,  \left [\psi_\mu\left(\partial_{\mu} D^*_\nu \bar{D}_{\nu}^{*} -D^*_\nu \partial_{\mu} \bar{D}_{\nu}^{*} \right ) 
                                                   +\left(\partial_{\mu} \psi_\nu D_{\nu}^{*}-\psi_\nu \partial_{\mu} D_{\nu}^{*}\right ) \bar{D}^*_\mu \right .  \nonumber \\ 
                                                & \left. +\ D^*_\mu \left(\psi_\nu \partial_{\mu} \bar{D}_{\nu}^{*}-\partial_{\mu} \psi_\nu \bar{D}_{\nu}^{*} \right ) \right ] \ , \\                                       
  \mathcal{L}_{\rho D D}          & = \ i g_{\rho D D}\left(D \vec{\tau}\, \partial_{\mu} \bar{D}-\partial_{\mu} D \vec{\tau}\, \bar{D}\right) \cdot \vec{\rho}_\mu \ ,
  \label{rhoDDcouple} \\              
  \mathcal{L}_{\rho D^{*}\! D}  & =\ g_{\rho D^* D} \, \epsilon_{\mu \nu \alpha \beta}\, \partial_{\mu} \rho_{\nu}\left(\partial_{\alpha} D_{\beta}^{*} \vec\tau\bar{D}
                                                   +  D \vec \tau \partial_{\alpha} \bar{D}_{\beta}^{*} \right ) \ , \\                              
  \mathcal{L}_{\rho D^*\! D^*}  & = \  i g_{\rho D^* D^*} \left [\left(\partial_{\mu} D^*_\nu \vec{\tau}\, \bar{D}_{\nu}^{*} -D^*_\nu \vec{\tau}\, \partial_{\mu} \bar{D}_{\nu}^{*}\right) \cdot \vec{\rho}_\mu
                                                   + \left ( D^*_\nu \vec{\tau} \cdot \partial_{\mu} \vec{\rho}_{\nu}-\partial_{\mu} D^*_\nu \vec{\tau} \cdot \vec{\rho}_{\nu}\right) \bar D^*_\mu \right. \nonumber  \\
                                                & \left. + \ D^*_\mu  \left(\vec{\tau} \cdot \vec{\rho}_\nu\, \partial_{\mu} \bar{D}_{\nu}^{*}-\vec{\tau} \cdot \partial_{\mu} \vec{\rho}_\nu \bar{D}_{\nu}^{*}\right)\right]  \ ,
  \label{rhoD*D*couple}                                                
\end{align}
\end{spreadlines}
In the above expressions, $\vec \pi$ and $\vec \rho$ denote the pion and rho meson isospin triplets, respectively, and $\vec \tau$ are  Pauli matrices, while $D\equiv (D^0,D^+)$ and 
$D^*\equiv (D^{*0}, D^{*+})$ describe the pseudoscalar and vector $D$-meson doublets.

The charge-specific coupling constants, $g_{PPV}$, $g_{PVV}$ and $g_{VVV}$, are functions of the universal SU$(4)_F$ couplings $g$ and $g_a$ defined with the Lagrangians
in Eqs~\eqref{phenLagrangian} and \eqref{anomLagrange} and their explicit relations are given in Table~1 of Ref.~\cite{Bracco:2011pg}. We merely remind that exact SU$(4)_F$ symmetry 
implies certain relationships between the couplings, amongst which we highlight:
\begin{align}
  g_{D^*D\pi} & =  g_{D^{*-} D^0 \pi^+ } =  g_{D^{*+} D^0 \pi^- } \, =  \sqrt{2}\, g_{D^{* \pm} D^{\mp} \pi^0 } = \sqrt{2}\, g_{D^{* 0} D^0 \pi^0 } = \tfrac{\hat g}{2\sqrt{2}}  \ ,  \\  
  g_{\psi D D}  & = g_{\psi D^{0} D^{0}}=g_{\psi D^{+} D^{-}}  \ ,  \quad  g_{\psi D D}  = g_{\psi D^{*} D^{*}}  =  \tfrac{\hat g}{\sqrt{6}} \ , \\
  g_{\rho D^* D} & = g_{\rho^{+} D^{0} D^{*-}} = g_{\rho^{-} \bar{D}^{0} D^{*+}}  = g_{ \rho^{+} D^{-} D^{* 0} } =  g_{\rho^{-} D^{+} \bar D^{* 0} } 
                            = \tfrac{1}{2\sqrt{2}}  \tfrac{\hat g_{a}^{2} N_{c}}{16 \pi^{2} f_{\pi}}   \ , \\
   g_{ D^* \! D\pi}  & = g_{\rho D D}  =  g_{\rho D^*\! D}  =  g_{\rho D^*\! D^*}  \ .                             
\end{align}
The SU$(3)_F$ relation $g_{\rho KK} = g_{\rho\pi \pi}/2$ is  preserved and generalized to:
\begin{equation}
     g_{\rho D D}  = g_{\rho K K} = \frac{g_{\rho \pi \pi}}{2} \  .
  \label{rhoDDrhopipirhoKK}   
\end{equation}
If one takes SU$(4)_F$ at face value, a single known coupling constant, for example $g_{D^{*+} D^0 \pi^+}$ which has been measured experimentally~\cite{CLEO:2001sxb}, suffices to infer all other
couplings. However, as we will see, in particular the last algebraic relation~\eqref{rhoDDrhopipirhoKK} strongly indicates that the SU$(4)_F$ breaking between couplings appears to be incompatible 
with the idea of a universal  coupling. We may thus ask: can $\hat g$ be unambiguously related to all the strong-interaction matrix elements mentioned above, $g_{D^* D\pi}$, $g_{D\rho D}$, etc.? 
Given that the rationale of building  Lagrangians for heavy mesons interactions with light mesons is nowadays extended to the SU$(5)_F$ flavor group~\cite{Zeminiani:2020aho}, our aim is to emphasize 
this shortcoming and to propose effective couplings from quark-degrees of freedom that will account for the flavor symmetry breaking effects without  resorting to additional terms in the $1/M_H$ expansion.


\section{Effective couplings and flavor SU(4)$_\textit{F}$ breaking  \label{sec3} }

Soon after the introduction of effective SU(4)$_F$ Lagrangians, such as the ones discussed in Section~\ref{sec2}, the study of interactions between open-charm mesons and nuclei to explore the 
possibility of charmed nuclear bound states became the object of intense activity~\cite{Sibirtsev:1999js, Mizutani:2006vq,Haidenbauer:2007jq,Molina:2009zeg,Haidenbauer:2010ch,Yamaguchi:2011xb}. 
Their application has gone beyond that field and include the reaction $D^{(*)}D^{(*)} \to \pi X(3872)$ which involves an intermediate triangle diagram with pseudoscalar and vector 
mesons~\cite{MartinezTorres:2014son}, the decay of a highly excited $D$ meson~\cite{Malabarba:2021gyq}, the $J/\psi$ regeneration in the hadronic gas phase following the cooling of a quark-gluon 
plasma~\cite{Abreu:2017cof}, the production of the $\psi(3770)$ resonance in $\bar pp \to \bar DD$~\cite{Haidenbauer:2015vra} or the thermal properties of pseudoscalar and vector charm 
mesons~\cite{Montana:2020vjg}, amongst others. At the same time, the effective couplings, quite naturally, have also been the object of interest~\cite{El-Bennich:2012hom,Belyaev:1994zk,
El-Bennich:2010uqs,El-Bennich:2011tme,El-Bennich:2016bno,Bracco:2001dj,Bracco:2007sg,OsorioRodrigues:2010fen,OsorioRodrigues:2013xvc,Cerqueira:2015vva,Wang:2007zm,Wang:2007mc,
Cui:2012wk,Can:2012tx,Khosravi:2013ad,Khosravi:2014rwa,Janbazi:2017mpb,Seyedhabashi:2019qjg,Fontoura:2017ujf,Ballon-Bayona:2017bwk,Aliev:2021cjt,deDivitiis:1998kj,Ohki:2008py,
Becirevic:2009yb,Becirevic:2009xp,Becirevic:2012pf,Flynn:2015xna,Bernardoni:2014kla,Detmold:2012ge,Braghin:2018drl,Braghin:2021hmr,Braghin:2020yri}.

\begin{figure}[t!]
\centering
  \includegraphics[scale=0.2]{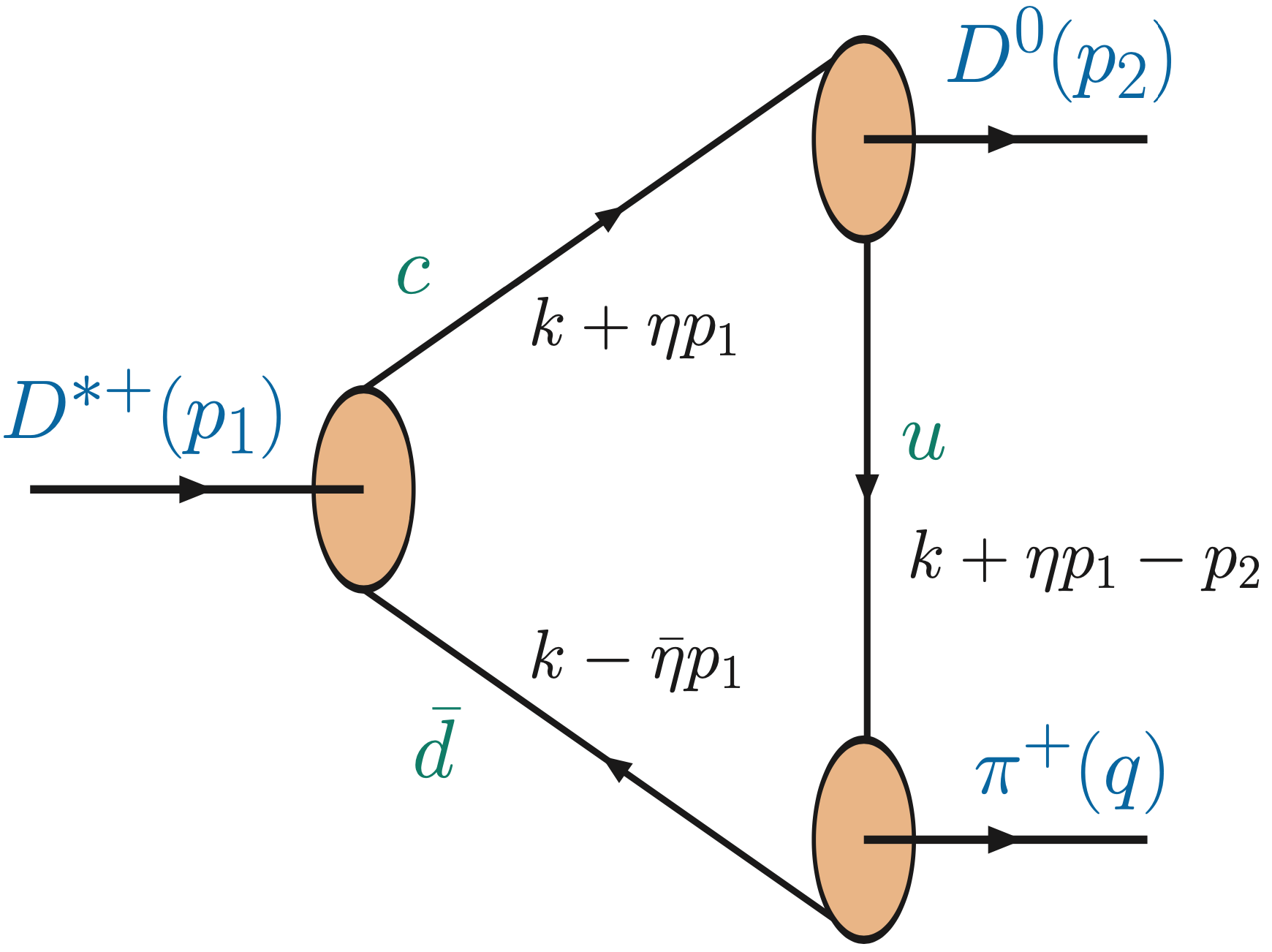}
  \caption{Diagrammatic representation of the impulse approximation to the $D^{*+} \to D^0\pi^+$ decay in Eq.~\eqref{H*Hpi}. The internal lines represent dressed quark propagators 
               and the filled ellipses are the meson's BSAs. In case of the soft-pion $B^{*+} \to B^0\pi^+$ amplitude the following substitutions are in order: $c \to \bar b$, $\bar d \to u$ and $u \to \bar d$.}
\label{fig1}
\end{figure}

We here summarize earlier results for the $D^*D\pi$, $D_s^*DK$, $B^*B\pi$, $B_s^*BK$, $\rho \pi\pi$, $\rho KK$, $\rho DD$, $\rho D^*D$  and $\rho D^*D^*$ couplings calculated in 
\emph{impulse approximation\/}, where the dressed heavy- and light- quark propagators were algebraic approximations to numerical solutions of the quark's Dyson-Schwinger equation 
(DSE)~\cite{Bashir:2012fs}, and the Bethe-Salpeter amplitude (BSA) for the $D$ and $B$ were modeled in absence of available solutions for asymmetric heavy-light bound states at that 
time~\cite{El-Bennich:2012hom,El-Bennich:2010uqs,El-Bennich:2011tme,El-Bennich:2016bno}. As already mentioned in Section~\ref{sec2}, the simplest coupling that gives access to 
$\hat g$ is found in the definition of the $D^{*+} \to D^0\pi^+$ decay amplitude:
\begin{equation}
   \langle  D(p_2) \pi(q) | D^*(p_1,\lambda)  \rangle \  :=  \   g_{D^*D\pi}\,  \bm{\epsilon}^{\lambda_{D^*}} (p_1) \cdot  q  \ .
\label{Hstarpicoupl}
\end{equation}
While here illustrated for a strong $D^*$ decay,  this matrix element generally defines the dimensionless coupling of a heavy-light vector meson $H^*$, characterized by a polarization state
$\lambda$, and a heavy-light pseudoscalar meson, $H$, to a pion with momentum $q = p_1 - p_2$. Hence, one may also use this amplitude for the unphysical process $B^*\to B\pi$
in the chiral $m_\pi \to 0$ limit which defines $g_{B^*B\pi}$. A consistent calculation within the same framework allows therefore for a quantitative estimate of the degree to which notions 
of heavy-quark symmetry are sensible in the charm sector. The diagrammatic representation of such a decay is depicted in Figure~\ref{fig1}. The decay $H^*\to H\pi$ amplitude in impulse 
approximation is given by,
\begin{equation}
  g_{H^*\!H\pi}\  \bm{\epsilon}^{\lambda_{H^*}}\!\! \cdot q   \ =  \ \operatorname{tr} \int\!\! \frac{d^4k}{(2\pi)^4} \  \bm{\epsilon}^{\lambda_{H^*}}\! \cdot \Gamma_{H^*}(k;p_1) \, S_Q(k_Q)\,
            \bar  \Gamma_H(k;-p_2)\,  S_u (k_u) \, \bar \Gamma_\pi(k;-q) \, S_d (k_d) \ ,
\label{H*Hpi}
\end{equation}
where the trace is over Dirac spinor and color indices,  $\bm{\epsilon}^{\lambda_{H^*}}$ is the vector-meson polarization and the momenta are $k_Q = k + \eta p_1$, 
$k_d = k - \bar \eta p_1$ and $k_u = k + \eta p_1 - p_2$  with the momentum partition parameters $\eta+\bar \eta = 1$. Moreover,  $S(k)$ and $\Gamma (k,p)$ are dressed quark 
propagators~\cite{Chang:2009ae} and BSAs, respectively, discussed in Section~\ref{sec4}.

The dimensionless coupling $g_{H^*H\pi}$  is related at leading order in the $1/M_H$ expansion to the strong meson coupling $\hat g$~\cite{Casalbuoni:1996pg}:
\begin{equation}
      g_{H^*\! H\pi} \ = \  2\,\frac{\sqrt{M_H M_{H^*}}}{f_\pi}\ \hat g \, .
  \label{ghat}    
\end{equation}
Using a constituent charm quark propagator and Eqs.~\eqref{H*Hpi} and \eqref{ghat} the difference in extracting $\hat g$ from either $g_{D^*\! D\pi}$ 
or $g_{B^*\!B\pi}$ is material~\cite{El-Bennich:2010uqs},
\begin{equation}
   g_{D^*\!D\pi}   =15.8^{+2.1}_{-1.0}\ \Longrightarrow \ \hat g_c = 0.53^{+0.07}_{-0.03}  \ ,   \qquad  g_{B^*\!B\pi}  = 30.0^{+3.2}_{-1.4} \ \Longrightarrow \ \hat g_b = 0.37^{+0.04}_{-0.02} \ ,
\label{eqcouple1}    
\end{equation}
where in order to distinguish between the extraction of the universal coupling from either $H^*H\pi$ vertex we label the quark flavor: $\hat g_c$ and $ \hat g_b$.
One can also employ a confining heavy-quark propagator of the algebraic form,
\begin{equation}
\label{SQ}
   S_Q (k) \ =  \ \frac{-i \gamma\cdot k + m_Q}{m_Q^2}\  \mathcal{F}(k^2/m_Q^2)\ ,  
\end{equation}
with the entire function ${\cal F}(x)= [1-\exp(-x)]/x$. Unlike the constituent-quark propagator $S_Q(k) = [ i\gamma\cdot k + m_Q ]^{-1}$, Eq.~\eqref{SQ} implements confinement
yet still produces a momentum independent heavy-quark mass-function. With this substitution for the heavy-quark propagator and with $ m_c = 1.32$~GeV and $ m_b = 4.65$~GeV
one obtains~\cite{El-Bennich:2012hom}:
\begin{equation}
   g_{D^*\!D\pi}   =18.7^{+2.5}_{-1.4} \ \Longrightarrow \ \hat g_c = 0.63^{+0.08}_{-0.05}  \ ,   \qquad  g_{B^*\!B\pi}  = 31.8^{+4.1}_{-2.8} \ \Longrightarrow \ \hat g_b = 0.39^{+0.05}_{-0.03} \ .
\label{eqcouple2}   
\end{equation}
Our value for $g_{D^*\!D\pi}$ is in good agreement with the coupling extracted by CLEO from the $D^*\to D\pi$ decay width: $g_{D^*\!D\pi}^\mathrm{exp.} = 17.9\pm1.9$~\cite{CLEO:2001sxb}.
A more recent analysis of the $D^*(2010)^+$ decay width by the BABAR collaboration~\cite{BaBar:2013zgp} yields a somewhat smaller coupling, $g_{D^*\!D\pi}^\mathrm{exp.} = 16.9\pm 0.14$, 
which still agrees with most theoretical calculations.  In Table~\ref{tab1} and \ref{tab2} we collect a set of representative values for the couplings obtained with our approach, 
QCD sum rules (QCDSR) and Lattice QCD.

\begin{table}[t!]
\centering
\begin{tabular}{ C{1cm} | C{2cm}| C{2cm} | C{1.5cm} | C{2cm}| C{1.8cm} | C{1.9cm} }
      & CLEO~\cite{CLEO:2001sxb} &  BABAR~\cite{BaBar:2013zgp}  & DSE~\cite{El-Bennich:2012hom} & QCDSR~\cite{Bracco:2011pg}  & Lattice~\cite{Becirevic:2009xp} & Lattice~\cite{Becirevic:2012pf}
       \\  \hline 
     $g_{D^{*} D \pi}$  & $17.9 \pm 1.92$ &  $16.9\pm 0.14$   & $18.7_{-1.4}^{+2.5}$ & $17.5 \pm 1.5$ & $ 20.0 \pm 2.0$ & $15.8\pm 0.8 $ 
       \\ \hline
      $\hat g_c$  &  $0.60 \pm 0.06$ &  $0.57 \pm0.006 $  &  $0.63^{+0.08}_{-0.05} $  & $0.59\pm 0.05$  & $0.67\pm 0.07$   &  $0.53\pm 0.03 $
\end{tabular}
\caption{Comparison of theoretical values for $g_{D^{*} D \pi}$ and $\hat g_c$ with those extracted from the experimental $D^*\to D\pi$ decay widths. Note that $\hat g_c$ is extracted at leading order
              via Eq.~\eqref{ghat}.    }
\label{tab1}
\end{table}

\begin{table}[t!]
\centering
\begin{tabular}{ C{1cm} | C{1.4cm}| C{1.9cm} | C{2.4cm} | C{1.8cm}| C{1.8cm} | C{1.8cm} }
&  DSE~\cite{El-Bennich:2012hom} & QCDSR~\cite{Belyaev:1994zk} & Lattice~\cite{Becirevic:2009yb} & Lattice~\cite{Flynn:2015xna} & Lattice~\cite{Bernardoni:2014kla} & Lattice~\cite{Detmold:2012ge}
       \\  \hline 
    $g_{B^{*} B\pi}$  & $31.8^{+4.1}_{-2.8} $ & $29.0 \pm 3.0 $ & ---  &  ---  &   ---    & ---  \\ \hline
    $ \hat g_b$   & $0.39^{+0.05}_{-0.03}$  &  $0.36\pm 0.04$  & $ 0.44\pm 0.03^{+0.07}_{-0.0}$  & $0.56\pm 0.08$  & $0.49\pm 0.03$ & $0.45\pm0.05$ 
\end{tabular}
\caption{Comparison of theoretical values for $g_{B^{*} B \pi}$  and $\hat g_b$ in the soft-pion limit.}
\label{tab2}
\end{table}

Turning our attention now to the $B^*B\pi$ coupling, we observe that $\hat g_b$ obtained with lattice QCD is uniformly above $0.4$, yet the result of Ref.~\cite{Flynn:2015xna} stands out by providing 
the sole value in agreement with the $D^*D\pi$ coupling. This is not generally the case and whether $\hat g$ is extracted from the $D^*D\pi$ or $B^*B\pi$ vertex in the DSE-BSE approach has 
significant consequences. Indeed, it is often interpreted as a sign that the leading term of a  $\Lambda_\mathrm{QCD}/m_c$ expansion may not be reliable. On the other hand, the numerical values 
obtained for either couplings in lattice QCD do not allow for a clear picture, i.e. whether $\hat g_c > \hat g_b$. 

We take advantage of this discussion to remind that the DSE-BSE calculations in Refs.~\cite{El-Bennich:2010uqs,El-Bennich:2012hom,El-Bennich:2011tme,El-Bennich:2016bno} employ algebraic 
expressions for the quark propagators  of the light quarks based on DSE solutions and a simple leading covariant BSA model for the $D$, $D^*$ $B$ and $B^*$ mesons. These simplifications 
were mainly imposed by the lack of proper BSAs for heavy-light mesons at that time~\cite{El-Bennich:2008dhc}. The shortcomings of  a too simplistic ladder-truncation of the BSE 
kernel~\cite{Rojas:2014aka,Rojas:2014tya,El-Bennich:2016qmb,Mojica:2017tvh} have meanwhile been overcome with distinct ans\"atze for the  quark-gluon vertex in the heavy and light 
sector~\cite{Serna:2017nlr,Serna:2018dwk,Serna:2020txe,Albino:2018ncl,Albino:2021rvj},  and a much improved, fully Poincar\'e-covariant calculation of $g_{D^*\!D\pi}$ and $g_{B^*\!B\pi}$ is 
underway~\cite{RCSilveira}.  

Of course, as in the case of the soft-pion $B^*B\to \pi$ decay, one may calculate other \emph{unphysical} strong decay amplitudes, for instance $B^*_s\to BK$ and $D^*_s\to DK$ from which
the effective couplings $g_{D_s^*DK}$ and $g_{B_s^*BK}$ can be extracted. This is because we can use Eq.~\eqref{H*Hpi} irrespective of the $q^2$-value and without resorting to extrapolations
from spacelike to timelike momenta or of the current-quark mass. The couplings obtained in Ref.~\cite{El-Bennich:2012hom} are:
\begin{equation}
   g_{D_s^* DK}   = 24.1^{+2.5}_{-1.6}  \ \  ,   \qquad  g_{B_s^* BK}  = 33.3_{-3.7}^{+4.0} \  .
\label{eqcouple3}   
\end{equation}
It is notable that the ratios, $g_{D_s^*DK} / g_{D^*\!D\pi}$ and $g_{B_s^*BK} / g_{B^*\!B\pi}$  are of comparable  magnitude as the decay-constant ratios $f_{D_s}/f_D$ and $f_{B_s}/f_B$, respectively,
indicating a typical $SU(3)_F$ flavor breaking pattern of the order of 20\%.

\begin{figure}[t]
\centering
\includegraphics[clip,width=0.52\textwidth]{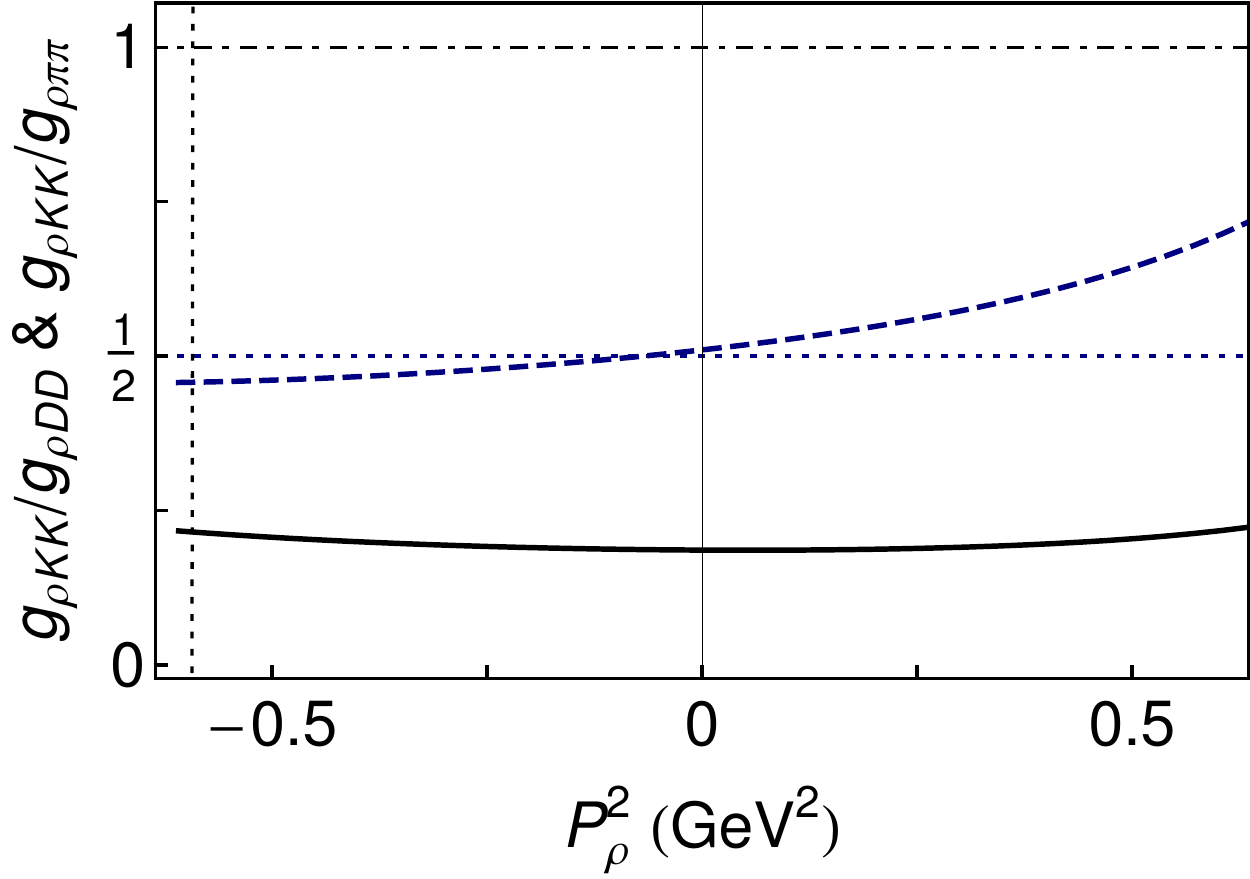}
\caption{Ratios of the couplings $g_{\rho KK}/g_{\rho DD}$ (solid curve) and $g_{\rho KK}/g_{\rho \pi\pi}$ (dashed curve). In case of exact flavor SU(4)$_F$ symmetry, 
these ratios take the values $1$ (dot-dashed line) and $1/2$ (dotted line), respectively. The vertical dotted line marks the on-shell point of the $\rho$-meson and in Euclidean metric 
$q^2>0$ is spacelike. Figure taken from Ref.~\cite{El-Bennich:2011tme}. }
\label{figcoupling}
\end{figure}

Further ratios that provide a measure for $SU(3)_F$ and $SU(4)_F$ breaking are based on the $\rho \pi\pi$, $\rho KK$, $\rho DD$, $\rho D^*D$ and $\rho D^*D^*$ couplings. The latter three 
couplings are those we came across in the effective Lagrangians~\eqref{rhoDDcouple}--\eqref{rhoD*D*couple}, whereas the first coupling relates directly to the strong decay $\rho \to \pi\pi$. 
While no physical processes are associated with the $\rho KK$, $\rho DD$, $\rho D^*D$ and $\rho D^*D^*$ couplings, they are commonly employed in defining $\rho$-mediated exchange interactions 
between a nucleon and kaons or $D$-mesons~\cite{Haidenbauer:2007jq,Haidenbauer:2010ch}. In these applications, the $\rho$-meson momentum is necessarily off-shell and spacelike, and 
couplings and form factors may be defined once one settles on a definition of the off-shell $\rho$-meson. At leading-order in a systematic, symmetry-preserving truncation scheme, in analogy
with the physical decay amplitude in Eq.~\eqref{H*Hpi}, we can express the $\rho DD$ matrix element as:
\begin{equation}
   g_{\rho DD} \  \bm{\epsilon}^{\lambda_\rho} \! \cdot p_1  \ = \  \operatorname{tr}  \int\! \frac{d^4k}{(2\pi)^4} \ \bar \Gamma_D (k, -p_1)  \, S_c(k_c) 
                 \, \bar  \Gamma_D(k, -p_2) \, S_f (k_f')\, \bm{\epsilon}^{\lambda_\rho}\! \cdot \Gamma_\rho(k, q) \, S_f (k_f)  \  ,
\label{mainamplitude}
\end{equation}
where we work in the isospin symmetric limit, so $f=u=d$. Momentum conservation requires $q=p_1+p_2$, $k_f =k+\eta q $,  $k_f' =k-\bar \eta q$ and $k_c = k+ \eta q -p_1$, where the 
relative-momentum partitioning parameters satisfy $\eta + \bar \eta = 1$ and $p_1^2 =p_2^2 = - M_D^2$. The integral expressions for $g_{\rho D^* D}$ and $g_{\rho D^* D^*}$, respectively 
Eqs.~\eqref{eq2} and \eqref{DstarrhoDstar}, are obtained with the substitutions $\Gamma_D (k;p)\, \to\, \bm{\epsilon}^{\lambda_{D^*}}\! \cdot \Gamma_{D^*} (k;p)$.

\begin{figure}[t!]
\centering
\includegraphics[width=0.48\textwidth]{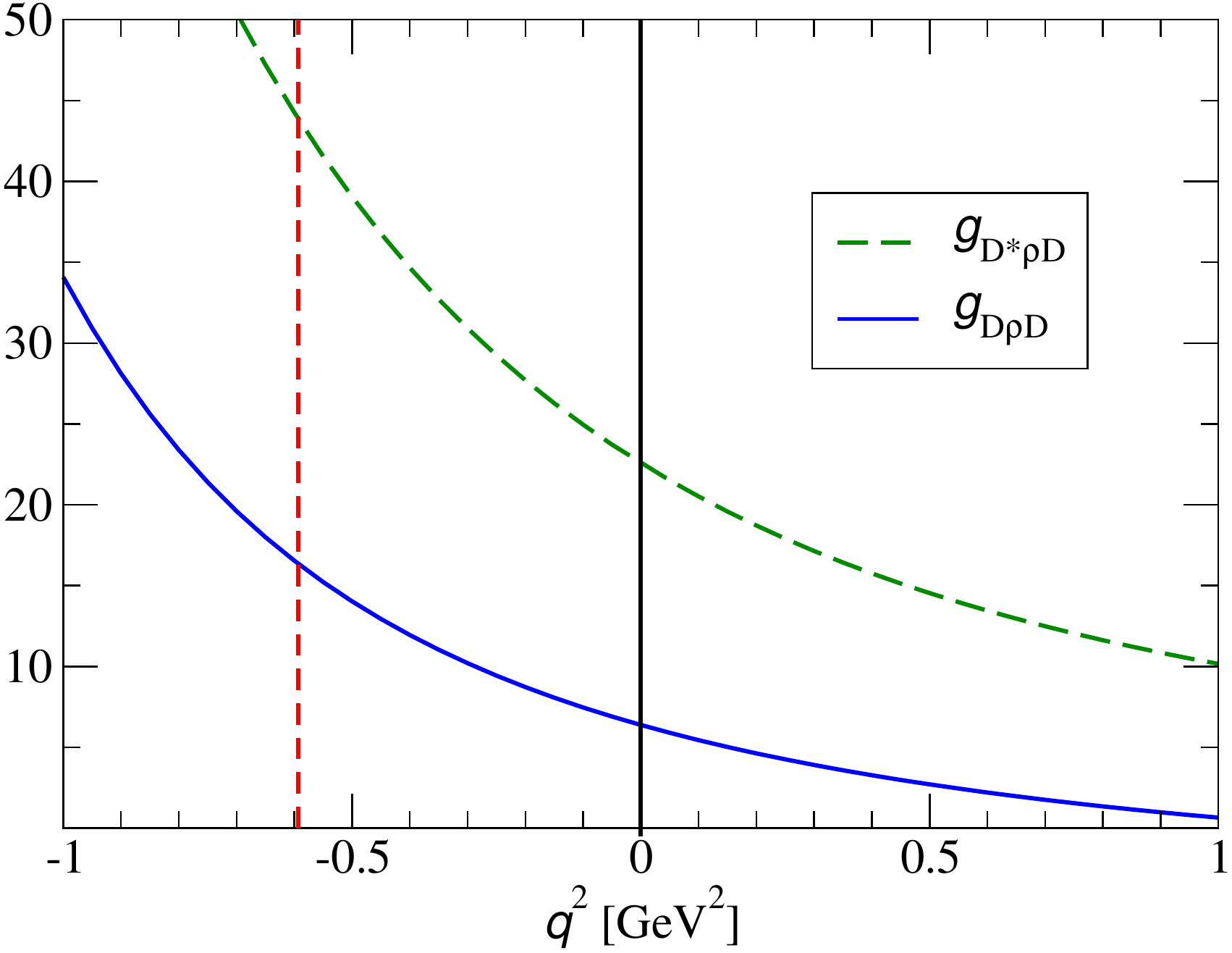}
\hfill \includegraphics[width=0.48\textwidth]{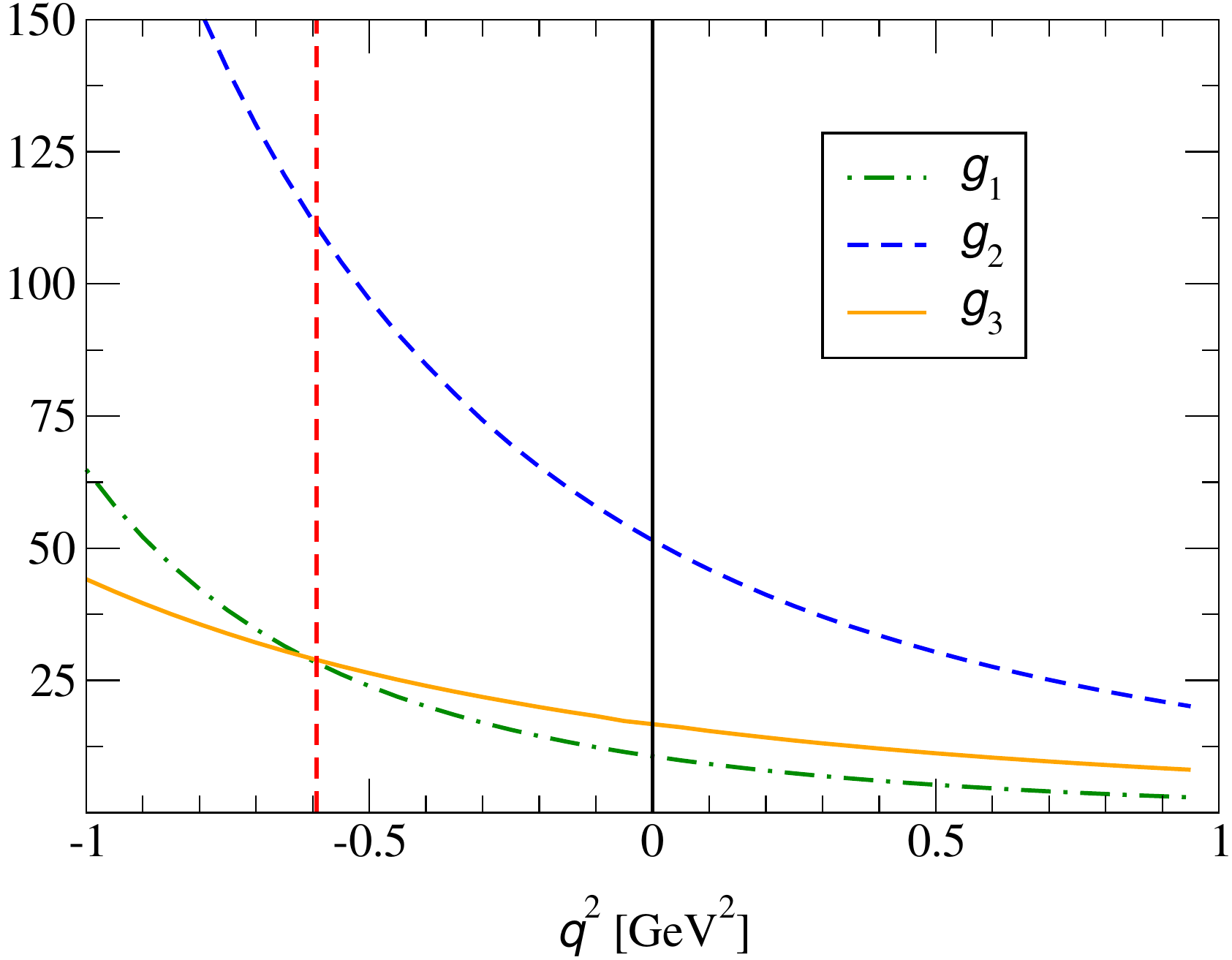}

\caption{   \emph{Left panel\/}:
The dimensionless couplings $g_{D \rho D}$  and $g_{D^*\! \rho D}$  as a function of the $\rho$-meson's four-momentum squared, with the $D$ and $D^*$ 
mesons on-shell.   Note that $g_{D^*\! \rho D}$ rather than $g_{D^*\! \rho D}/m_{D^*}$ is plotted and $q^2 > 0$ is spacelike in Euclidean metric.
\emph{Right panel\/}: The three-vector couplings defined in Eq.~\eqref{DstarrhoDstar} with the abbreviation $g_{D^*\! \rho D^*}^i \equiv  g_i,\, i=1,2,3 $, where the $D^*$-mesons are on-shell.
In both panels the dashed vertical line denotes the $\rho$-meson on-shell point. Figures taken from Ref.~\cite{El-Bennich:2016bno}.  }
\label{Fig4}
\end{figure}

As we highlighted in Eq.~\eqref{rhoDDrhopipirhoKK}, SU(4)$_F$ symmetry implies some stringent relations between the couplings. As can be read from Figure~\ref{figcoupling}, the equality 
$g_{ \rho KK} = g_{\rho \pi\pi}/2$ provides a fair approximation on the domain $P^2\in [-m_\rho^2,m_\rho^2]$  where the deviation ranges from $-10\%$ to 40\% and describes again the 
typical SU(3)$_F$ breaking pattern. When it comes to the relation $g_{\rho DD} = g_{\rho KK}$, the striking discrepancy between what flavor symmetry dictates and what is found in a covariant 
calculation based on quark degrees of freedom is of the order of 360\% to 440\%. We note that a lattice QCD  calculation~\cite{Can:2012tx} finds $g_{\rho DD}(0) = 4.84(34)$ and 
$g_{\rho D^*D^*}(0) = 5.94(56)$, values that compare well with our coupling in Figure~\ref{Fig4} at zero-recoil momentum:  $g_{\rho DD}(0) \simeq 6.3$.

As we continue with the couplings that involve a more complicated spin structure, we start with the amplitude that describes the transition of on-shell $D^*$ to $D$ mesons emitting an off-shell 
$\rho$-meson or equivalently, the unphysical process of an off-shell $\rho$-meson decaying into a $D^*D$ pair.  This defines the $g_{\rho D^*D}$ couplings as follows:
\begin{equation}
   g_{\rho D^* D} \, \frac{1}{M_{D^*}} \, \varepsilon_{\alpha\beta\mu\nu}\,  \bm{\epsilon}_\alpha^{\lambda_{D^*}} \bm{\epsilon}_\beta^{\lambda_\rho} \ p_{1\mu}\, p_{2\nu} 
                            \  := \  \langle  D^*(p_2,\lambda_{D^*}) | \, \rho (q,\lambda_\rho)\,  | D (p_1 ) \rangle\,.   
        \label{eq2}
\end{equation}
The three vector-meson vertex $\rho D^* D^*$, with two mesons on-shell,  introduces additional complexity and requires a minimal set of three independent couplings analogous 
to the electromagnetic form factors of vector mesons~\cite{Hawes:1998bz, Bhagwat:2006pu}:
\begin{align}
   \langle  D^*(p_2\, ,\lambda_{D^*}) | \, \rho (q,\lambda_\rho)\,  | D^* (p_1\, ,\lambda_{D^*}) \rangle  
    & = \  - \sum_{i=1}^{3} T_{\mu\rho\sigma}^{\,i} (p, q) \; g^i_{D^*\!\rho D^*}(q^2) \,  \bm{\epsilon}_\mu^{\lambda_{\rho} } \bm{\epsilon}_\rho^{\lambda_{D^*}} \bm{\epsilon}_\sigma^{\lambda_{D^*} } 
    \nonumber \\
   & = \  \Lambda_{\mu\rho\sigma}(p,q)\, \bm{\epsilon}_\mu^{\lambda_{\rho} }\bm{\epsilon}_\rho^{\lambda_{D^*}}  \bm{\epsilon}_\sigma^{\lambda_{D^*}} \ ,  
\label{DstarrhoDstar}
\end{align}
%
%
with the tensor structures:
\begin{eqnarray}
  T_{\mu\rho\sigma}^{\,1}(p,q) & = & 2\, p_\mu \, \mathcal{ P}_{\rho\gamma}^T(p_1)\,  T_{\gamma\sigma} (p_2) \ ,
 \label{Eq:HPT1}  \\  [10pt]
  T_{\mu\rho\sigma}^{\,2}(p,q) & = & - \left[q_\rho - p_{1 \rho} \, \frac{q^2}{2\,  m_{D^*}^2}\right]  T_{\mu\sigma} (p_2) 
                   + \left[q_\sigma + p_{2\sigma} \, \frac{q^2}{2\, M_{D^*}^2}\right]  T_{\mu\rho} (p_1) \ ,
 \label{Eq:HPT2}  \\  [10pt]
 T_{\mu\rho\sigma}^{\, 3}(p,q) &=& \frac{p_\mu}{M_{D^*}^2} \, \left[q_\rho - p_{1 \rho}\, \frac{q^2}{2\,  M_{D^*}^2}\right]  \left[q_\sigma + p_{2 \sigma}\, \frac{q^2}{2\, M_{D^*}^2}\right] \ .
  \label{Eq:HPT3}
\end{eqnarray}
In Eqs.~\eqref{Eq:HPT1} to \eqref{Eq:HPT3} we introduce the four-momentum $p$ via $p_1 =p - \tfrac{1}{2} q$ and $p_2 =p + \tfrac{1}{2} q$, $p_1^2 =p_2^2 = -M_{D^{(*)}}^2 \,$
and $T_{\alpha\beta} ( p) =  \delta_{\alpha\beta} - p_\alpha p _\beta/p^2 $ is the transverse projection operator.  With this decomposition of the tensor structures all couplings 
are positive, $g^i_{D^*\!\rho D^*}(q^2) \geq 0,\ i=1,2,3$, and the matrix element  $\langle  D^*(p_2\, ,\lambda_{D^*}) | \, \rho (q,\lambda_\rho)\,  | D^* (p_1\, ,\lambda_{D^*}) \rangle$ satisfies
transversality:
\begin{eqnarray}
    p_{2\rho}\, \Lambda_{\mu\rho\sigma}(p,q) & = &  0  \ , \\
    p_{1\sigma}\, \Lambda_{\mu\rho\sigma}(p, q) & = &  0 \   , \\
    q_\mu\, \Lambda_{\mu\rho\sigma}(p,q) & = &   0 \ .
\end{eqnarray}
The corresponding couplings were obtained in Ref.~\cite{El-Bennich:2016bno} and are plotted as functions of the off-shell $\rho$-momentum in Figure~\ref{Fig4}. The dimensionless couplings $g_{\rho D D}$ 
and $g_{\rho D^*D}$ are smooth and monotonically decreasing as $q^2$ increases away from the on-shell point $q^2=-m_\rho^2$. On the average, in the domain $q^2 \in [-m_\rho^2, m_\rho^2]$, 
one  observes $g_{\rho D^*D} \gtrsim 3\,g_{\rho DD}$, which can  be attributed to differences in the BSA normalizations of the $D$ and $D^*$ and therefore indirectly to $f_{D^\ast} > f_D$.  
This yields the hierarchy relation:
\begin{equation}
     g_{\rho DD} (0) \approx 6.4  < g_{ D^*\!D \pi } \approx  17  <  g_{ \rho D^*\!D}(0) \approx 23 \ .
\end{equation}

The $\rho D^*\!D^*$ couplings are also smooth and monotonically decreasing functions of the $\rho$-momentum. However, there are remarkable quantitative differences between their magnitudes and 
damping rates. Averaging over the interval $q^2 \in [-m_\rho^2 ,m_\rho^2]$, one finds:
\begin{equation}
    \bar g_2 \big (q^2 \big ) \ \approx 3  \bar g_3  \big  (q^2 \big ) \  \approx \ 5  \bar g_1 \big (q^2 \big )  \ .
\end{equation}
Such relative strengths are of the same magnitude as those found in the $\rho$-meson elastic form factor. We also note that $\bar g_3(s) \approx 0.7\,\bar g_{\rho D^*\!D} (s)$, in other words one of 
the $ \rho D^*\!D^*$ couplings is of similar strength than the $\rho D^*\!D$ coupling.

\begin{figure}[t!]
\vspace*{-3cm}
\centering
\includegraphics[width=0.7\textwidth]{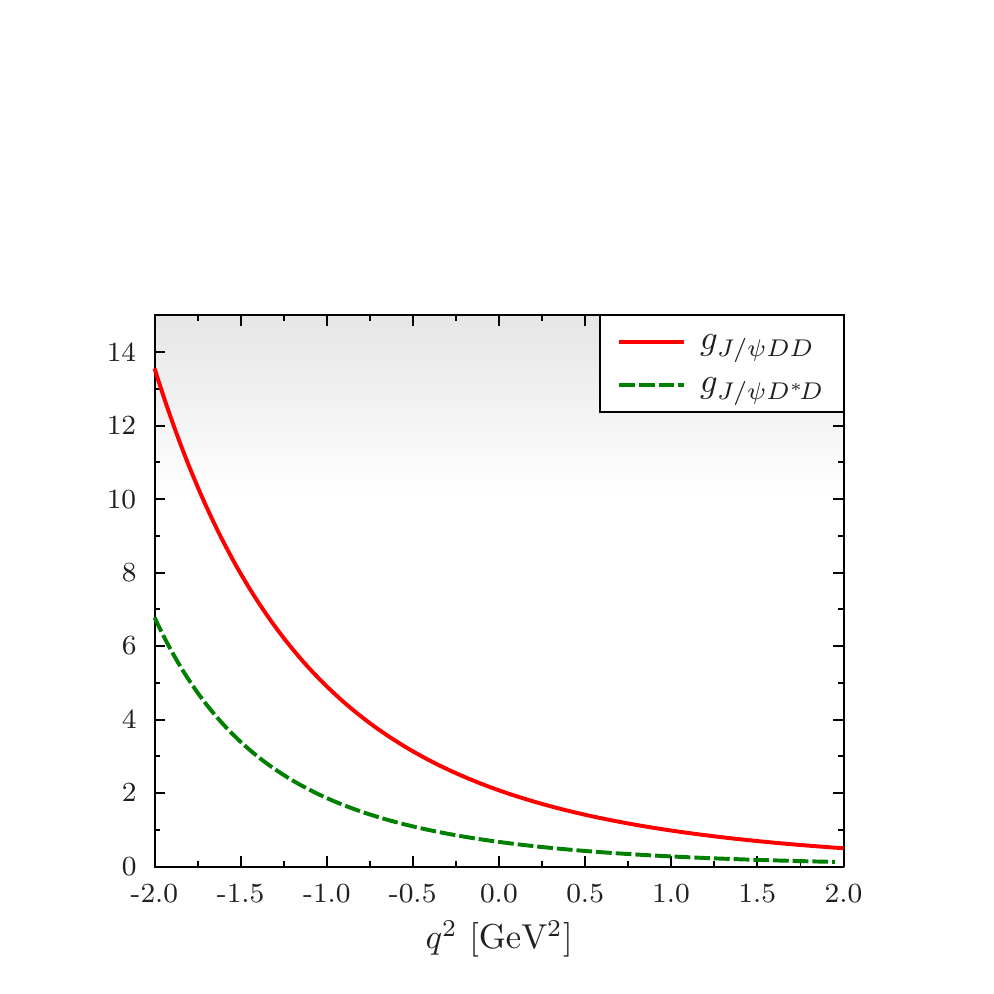}
\caption{The couplings $ g_{J/\psi DD} $ and $g_{J/\psi D^*\!D}$ as functions of the off-shell momenta square, $p_D^2 = p_{D^*}^2 = q^2$, and with the $J/\psi$ meson on-shell.}
\label{Jpsicoupling}
\end{figure}

In concluding this section, we extend the calculations of charmed couplings to the $J/\psi$ charmonium and present novel results within the same framework. 
In Figure~\ref{Jpsicoupling}, the $J/\psi D\bar D$ and  $J/\psi D^*\!\bar D$ couplings, defined by $\langle D(p_1) D(p_2) | J/\psi(p, \lambda_{J/\psi}) \rangle$ and  
$\langle D^*(p_1,\lambda_{D^*}) D(p_2) | J/\psi(p,\lambda_{J/\psi} ) \rangle$, are plotted, where the $J/\psi$ is on-shell whereas the $D$ mesons are \emph{both\/} symmetrically off-shell: 
$p_1^2 = p^2_2 = q^2$. Note that this differs from the convention in Ref.~\cite{Bracco:2011pg} where either the $J/\psi$ or one of the $D$ mesons are off-shell. This choice is motivated by the 
observation that the couplings are commonly required for space-like momenta of the $D$ and $D^*$ mesons in loop diagrams that describe $J/\psi \to D^{(*)}\!\bar D \to J/\psi$. Due to the 
simplified ansatz for the mesons's BSA based on Gaussian-type functions~\cite{El-Bennich:2016bno} and the use of two charm propagators defined in Eq.~\eqref{SQ}, these form factors 
grow rapidly for larger time-like momenta and we only plot them up to  $q^2 =-2$~GeV$^2$, which is below the on-shell point of the $D$ meson. These couplings are often quoted at 
$q^2 =0$~GeV$^2$,  where we find,
\begin{equation}
   g_{J/\psi DD} (0)  \simeq  2.2   \ \  ,   \qquad  g_{J/\psi D^*\!D} (0) \simeq 0.7~\mathrm{GeV}^{-1} \  ,
\end{equation}
remembering that both $D$-mesons are evaluated off-shell at zero-momentum. As for the $\rho DD$ couplings, we expect a more realistic behavior as a function of $q^2$ employing the full 
Poincar\'e covariant computation of the BSA  for the  $D$, $D^*$ and $J/\psi$ mesons~\cite{RCSilveira} rather than model wave functions.


\section{Poincar\'e covariance in bound-state description of heavy mesons  \label{sec4}}

In the light-meson sector it is crucial to satisfy chiral symmetry and its breaking pattern to ensures the pion is massless in the chiral limit. The axialvector Ward-Green-Takahashi identity (axWTI) 
describes the properties of the divergence of the vertex and expresses chiral symmetry and its breaking pattern:
\begin{equation}
  P_{\mu} \Gamma_{5 \mu}^{f g}(k ; P)  =   S_f^{-1} \left (  k_\eta \right) i \gamma_5 +  i \gamma_5 S_g^{-1}\left(k_{\bar \eta}  \right  )
                                                                - i  \left[ m_f+m_g \right ] \Gamma_5^{f g}(k ; P) \ .
 \label{axWTI}                                                               
\end{equation}
In this axWTI, $\Gamma_{5 \mu}^{f g}(k ; P)$ and $ \Gamma_5^{f g}(k ; P)$  are the color-singlet axialvector and pseudoscalar vertices for two quark flavors, $f$ and $g$, in 
Eqs.~\eqref{Gamma5mu} and  \eqref{Gamma5}. They are solutions of a BSE with $\gamma_5\gamma_\mu$ or $\gamma_5$ inhomogeneity, respectively. The total four-momentum of the 
meson satisfies $P^2 = - M_P^2$ and the quark momenta are defined as, $k_\eta = k + \eta P$ and $k_{\bar \eta} = k - \bar \eta P$. The inverse of the dressed quark propagator,  
$S_f^{-1} = \left [ i \gamma \cdot p \,A_f (p^2) + B_f ( p^2 ) \right  ]^{-1}$, is fully described by vector and scalar pieces whose respective dressing functions, $A_f (p^2)$ and $ B_f ( p^2 )$, 
are solutions of a DSE for a given flavor $f$ whose derivation is discussed in detail in Ref.~\cite{Holl:2006ni}, for example.  We  remind that the DSE is nothing else but the 
nonperturbative equation of motion of a particle in a relativistic quantum field theory. In QCD, this DSE is given by the integral equation,
\begin{equation}
   S^{-1}_f (p)  =  \, Z_2^f\,  i\, \gamma \cdot  p + Z_4^f\,  m_f (\mu)  \
                    +  \ Z_1^f  g^2  \! \int^\Lambda \! \frac{d^4k}{(2\pi)^4}  \, D^{ab}_{\mu\nu} (q) \frac{\lambda^a}{2} \gamma_\mu S_f(k) \, \Gamma^b_{\nu,f}  (k,p) \ ,
\label{QuarkDSE}
\end{equation}
where the integral describes the infinite tower of the quark's gluon dressing and $\Lambda \gg \mu$ is a Poincar\'e-invariant regularization scale that can be taken to infinity. 
In Eq.~\eqref{QuarkDSE}, $m_f(\mu)$ is the renormalized current-quark mass which is related to the bare mass in the QCD Lagrangian via $Z_4^f  (\mu,\Lambda )\, m_f(\mu)  =
 Z_2^f  (\mu,\Lambda ) \, m_f^{\rm bm} (\Lambda)$, while $Z_1^f(\mu,\Lambda)$, $Z_2^f(\mu,\Lambda)$ and $Z_4^f  (\mu,\Lambda )$ are respectively  flavor-dependent vertex, 
 wave function and mass renormalization constants. Moreover, $D_{\mu\nu}(q)$ is the dressed-gluon propagator  and $\Gamma^a_{\mu,f} (k,p) = \frac{1}{2}\,\lambda^a \Gamma_{\mu,f} (k,p) $ 
 is the quark-gluon vertex, where $\lambda^a$ are the SU(3) color matrices and $q=k-p$. The most general covariant form of the quark propagator is written in terms of scalar and vector 
 contributions with  $Z_f (p^2 ) = 1/A_f (p^2)$ and $M_f ( p^2 ) = B_f ( p^2 )/A_f ( p^2 )$:
\begin{equation}
\label{DEsol}
   S_f (p)  \  = \  \frac{ 1 }{i \gamma \cdot p \, A_f (p^2)   + B_f ( p^2 ) }  \ = \   \frac{ Z_f (p^2 ) }{  i \gamma \cdot p + M_f ( p^2 ) }  \ .
\end{equation}

Suppose we take a timid step beyond the abundantly employed rainbow-ladder (RL) truncation of the DSE and BSE, namely $\Gamma_{\mu,f} (k,p) \equiv \gamma_\mu$, motivated by the observation 
that the Abelian WTI it implies for the quark-gluon vertex\footnote{\,Instead of the correct Slavnov-Taylor identity.}, 
\begin{equation}
  i q \cdot  \gamma   \ =   \   i  k \cdot  \gamma \, A_f (k^2 )   - \, i  p \cdot  \gamma  \, A_f (p^2 )  
                                + \, B_f ( k^2 )   -  \, B_f ( p^2 )  \ ,
\end{equation} 
is only valid when $B_f ( k^2 ) \simeq B_f ( p^2) $ and $A_f ( k^2 ) \simeq A_f ( p^2)\simeq 1 $  over a large domain, $k^2, p^2 > 0$, of momentum squared. This is only true in the infinitely-heavy 
quark limit, but does not apply to the charm quark for which dressing effects are important~\cite{El-Bennich:2012hom}. Clearly, the vertex must exhibit some sort of flavor dependence.  Let us therefore
slightly modify the RL kernel of the DSE:
\begin{equation}
    Z_1^f g^2  D_{\mu\nu} (q)\, \Gamma_{\nu, f} (k, p) \ = \ \left ( Z^f_2 \right ) ^{\!2} \mathcal{G}_f (q^2) D_{\mu\nu}^\mathrm{free} (q) \frac{\lambda^a }{2} \gamma_\nu \ .
\label{RLtrunc}
\end{equation}
In here, the free gluon propagator in Landau gauge, 
\begin{equation}
    D_{\mu\nu}^\mathrm{free} (q) \ = \  \delta^{a b}\left(\delta_{\mu \nu}-\frac{q_{\mu} q_{\nu}}{q^{2}}\right ) \! \frac{1}{q^2} \ ,
 \label{freegluon}   
\end{equation}    
and bare vertex $\gamma_\nu$ are multiplied by an effective interaction model $\mathcal{G}_f (q^2)$ for the product of the gluon and vertex dressing which is now \emph{flavor dependent},
unlike in common RL models~\cite{Rojas:2014aka,Mojica:2017tvh}. 

Whatever modifications we apply to the DSE kernel must have repercussions in the BSE.  To this end, the effect of the ansatz~\eqref{RLtrunc} that preserves the axWTI~\eqref{axWTI} can 
be studied by inserting the  DSE~\eqref{QuarkDSE} as well as the  axialvector and pseudoscalar vertices given by,
\begin{eqnarray}
  \Gamma_{5 \mu}^{f g}  (k ; P) & = & Z_2^f   \gamma_5 \gamma_{\mu}   +  \int^\Lambda \! \!   \frac{d^4q}{(2\pi)^4}  \,  K_{f g} (q, k ; P) \, S_f ( q_\eta ) \Gamma_{5 \mu}^{f g}(q ; P) S_g (q_{\bar \eta} )  \,  ,  
\label{Gamma5mu}         \\
  \Gamma_5^{f g}  (k ; P)   & =  &  Z_4^f   \gamma_5  +   \int^\Lambda  \!\!  \frac{d^4q}{(2\pi)^4}  \, K_{f g} (q, k ; P) \,  S_f ( q_\eta ) \Gamma_5^{f g} (q ; P) S_g (q_{\bar \eta} )  \, ,                                   
\label{Gamma5}  
\end{eqnarray}
into the axWTI.~\eqref{axWTI}  which leads to the relation ($l=k-q$), 
\begin{align}
\label{WTImod} 
  &   \int^\Lambda \!   \frac{d^4q}{(2\pi)^4}   \,  K_{f g}(q, k ; P)   \big [ S_f  (q_{\eta} )  \gamma_5 +  \gamma_5  S_g  (q_{\bar \eta }) \big  ]  =   \nonumber  \\   
       -\  & \int^\Lambda \!  \frac{d^4q}{(2\pi)^4} \,
      \gamma_{\mu} \big [ \Delta_{\mu \nu}^f (l)    S_f  (q_{\eta} )  \gamma_5   +   \gamma_5  \Delta_{\mu \nu}^{g}(l)  S_g (q_{\bar \eta } )  \big ] \gamma_{\nu} \,  ,
\end{align}
In Eqs.~\eqref{Gamma5mu} and  \eqref{Gamma5}, $K_{f g}(q, k ; P)$ is the fully-amputated quark-anti-quark scattering kernel suppressing Dirac and color indices, $q_\eta = q + \eta P$ and 
$q_{\bar \eta} = q - \bar \eta P$  are the quark and antiquark momenta and we have defined:
\begin{equation}
  \Delta_{\mu \nu}^f (l) = \frac{4}{3}\, \big (Z_2^f \big )^{\! 2}  \, \mathcal{G}_f (l^2)  \left ( \delta_{\mu \nu} - \frac{l_\mu l_\nu}{l^2}\right ) \frac{1}{l^2} \ .
\end{equation} 
Comparing both sides of Eq.~\eqref{WTImod} one readily acknowledges that for $\Delta_{\mu \nu}^f (l) =  \Delta_{\mu \nu}^g (l)$ the identity Eq.~\eqref{WTImod} is satisfied 
by the usual RL kernel,
\begin{equation}
   K(q,k;P) \  =  \  - Z_2^2 \, \mathcal{G}\left(l^{2}\right)  D_{\mu \nu}^{\mathrm{free}}(l)\,  \gamma_{\mu} \frac{\lambda^a}{2} \gamma_{\nu} \frac{\lambda^a}{2} \ .
\end{equation}
On the other hand, with a flavor-dependent interaction $\mathcal{G}_f (l^2) $ the kernel $ K_{f g}(q, k ; P) $ on the  left-hand side of Eq.~\eqref{WTImod} must somehow produce a
flavored average of interactions. 

Indeed, a consistent ansatz~\cite{Qin:2019oar} for $K_{f g}(q, k ; P)$ that satisfies Eq.~\eqref{WTImod} behaves for large momenta $q^2$ as, 
\begin{equation}
   K_{fg}  \ \sim \  -\, \gamma_\mu  \left(  \frac{\Delta_{\mu \nu}^f + \Delta_{\mu \nu}^g}{2}  \right ) \gamma_\nu  \ ,
\end{equation}
whereas in the infrared limit this becomes,
\begin{equation}
  K_{fg} \ \sim \  -\, \gamma_\mu  \left (  \frac{\Delta_{\mu \nu}^f \sigma_s^f (0) + \Delta_{\mu \nu}^g \sigma_s^g (0) }{\sigma_s^f (0) + \sigma_s^g (0) }  \right )    \gamma_\nu \, .
\end{equation}
Both limits describe and average of interaction functions, in the latter case weighted with flavored quark-dressing functions.

Given these considerations, we modified the RL truncation by introducing a flavor dependence in the effective DSE and BSE vertices~\cite{Serna:2020txe} which leads
to a different treatment of the light and heavy quarks. For the former, the vertex dressing is of significant magnitude, while for the latter it amounts to describing the quark-gluon
interaction with all but a bare vertex; see Fig.~1 in Ref.~\cite{Serna:2020txe} for a comparison of the interaction strength of light ($u,d,s$) and heavy ($c,b$) quarks.  
The BSE kernel is thus written as, 
\begin{equation}
  \label{RLkernel}
    K_{fg}(k,q;P)  \ = \ -  \mathcal{Z}_2^2  \  \frac{\mathcal{G}_{fg}  (l^2)}{l^2 } \, \frac{\lambda^a }{2} \gamma_\nu \frac{\lambda^a }{2} \gamma_\nu \ ,
\end{equation}
where the wave-function renormalization constants obtained for both quarks with the DSE~\eqref{QuarkDSE} is combined into a single one: $ \mathcal{Z}_2 (\mu,\Lambda) =\sqrt{ Z_2^f  Z_2^g}$.
For the averaged interaction we employ the ansatz,
\begin{equation}
    \frac{ \mathcal{G}_{fg}  (l^2) }{ l^2 } \ = \ \mathcal{G}_{fg}^\mathrm{ IR}(l^2) +  4\pi \tilde\alpha_\mathrm{PT}(l^2) \ ,
\end{equation}
in which the low-momentum domain is described by the Gaussian, infrared-finite support,
\begin{equation}
\label{BSEflavor}
 \mathcal{G}_{fg}^\mathrm{IR} (l^2) \ = \  \frac{8\pi^2}{(\omega_f\omega_g)^2} \sqrt{D_f\,D_g }\ e^{-l^2/(\omega_f\omega_g)} \ ,
\end{equation}
and the perturbative tail is given by the usual expression~\cite{Bashir:2012fs},
\begin{equation}
  4\pi \tilde\alpha_\mathrm{PT}(q^2)  \ = \ \frac{8\pi^2  \gamma_m \mathcal{F}(q^2)}{ \ln \left [  \tau +\left (1 + q^2/\Lambda^2_\textrm{\tiny QCD} \right )^{\!2} \right ] }  \ , 
\end{equation}
with $\gamma_m=12/(33-2N_f)$ being the anomalous dimension, $N_f$ is the active flavor number, $\Lambda_\textrm{\tiny QCD}=0.234$~GeV, $\tau=e^2-1$, 
$\mathcal{F}(q^2)=[1-\exp(-q^2/4m^2_t)]/q^2$ and $m_t=0.5$~GeV.  

With this ansatz in the homogeneous BSE for a meson $M=P,V$ becomes:
\begin{equation}
\label{BSE}
  \Gamma^{fg}_M (k,P) \!=\! \int^\Lambda\! \! \!\frac{d^4q}{(2\pi)^4 } \, K_{fg} (k,q;P)  S_f (q_\eta)\, \Gamma^{fg}_M (q,P)\, S_g(q_{\bar \eta} )\  .
\end{equation}
The solutions of the BSA for a pseudoscalar meson with quantum numbers $J^{PC} = 0^{-+}$ can most generally be decomposed into four Lorentz covariants made from the 
Dirac matrices $\gamma_\mu$, the relative momentum $k_\mu$ and the total momentum $P_\mu$. For the pseudoscalar mesons, we employ the tensor structures that are not 
orthogonal with respect to the Dirac trace,
\begin{equation}
   \Gamma^{fg}_P (k,P) =\   \gamma_5   \Big [ i E^{fg}_P (k,P) +  \gamma \cdot P\,  F^{fg}_P (k,P) \  +\  \gamma \cdot k\;  k \cdot P \,  G^{fg}_P (k,P) 
     + \sigma_{\mu\nu} k_\mu P_\nu \, H^{fg}_P (k,P) \Big ] \ ,
 \label{PS-BSA}                                      
\end{equation}     
where $E^{fg}_P (k,P)$, $F^{fg}_P (k,P)$, $G^{fg}_P (k,P)$ and $H^{fg}_P (k,P)$ are Lorentz-invariant amplitudes.

Likewise, the most general Poincar\'e-invariant form of the BSE solution for the vector vertex $\Gamma_{V\mu}$ in the $J^{PC} = 1^{--}$ vector channel is decomposed into eight
Lorentz covariants:
\begin{align}    
   \Gamma^{fg}_{V\mu} (k, P) \ = \  \sum^{8}_{\alpha=1} \, T^{\alpha}_{\mu}(k, P)\, \mathcal{F}_{\alpha}^{fg} (k, P ) \ .
\label{vectordiracbase}                            
\end{align}
In Eq.~\eqref{vectordiracbase}, $\mathcal{F}_\alpha (k,P)$ are Lorentz invariant amplitudes and $T^\alpha_\mu (k,P)$ is the orthogonal basis with respect to the Dirac trace:
\begin{eqnarray}
T^1_{\mu}(k,P) &=&  i\gamma^T_\mu \ , 
  \\[0.3true cm]
T^2_{\mu}(k,P) &=& i[3k^T_\mu(\gamma\cdot k^T)-\gamma^T_\mu(k^T)^2] \ ,
  \\[0.3true cm]
T^3_{\mu }(k,P) &= & i(k\cdot P) k^T_\mu \gamma\cdot P \ , 
  \\[0.3true cm]
T^4_{\mu}(k,P) &=& i[\gamma^T_\mu \gamma\cdot P(\gamma\cdot k^T)+k^T_\mu \gamma\cdot P] \ ,
  \\[0.3true cm]
T^5_{\mu}(k,P)&=&k^T_\mu \ , 
   \\[0.3true cm]
T^6_{\mu}(k,P)&=& (k\cdot P)[\gamma^T_\mu(\gamma\cdot k^T)-(\gamma\cdot k^T)\gamma^T_\mu] \ ,
  \\[0.3true cm]
T^7_{\mu}(k,P)&=&\gamma^T_\mu \gamma\cdot P - \gamma\cdot P \gamma^T_\nu -2 T_\mu^8(k,P) \ ,
  \\[0.3true cm]
T_{\mu}^8(k,P) &=& \hat k^T_\mu(\gamma\cdot \hat k^T) \gamma\cdot P \ .
\end{eqnarray}
The transverse projections are  $V^{T}_{\mu}= V_{\mu} - P_{\mu}(P\cdot V)/ P^2$ with $P\cdot V^T =0$ for any four-vector $V_\mu$ and $\hat k^T \cdot\hat k^T=1$.
Moreover, the BSA of both, pseudoscalar and vector mesons, also depend on the angle $ z_k = k\cdot P/|k||P|$ which is commonly exploited to expand the eigenfunctions
$\mathcal{F}_{\alpha}^{fg} (k, z_k, P)$ into Chebyshev polynomials. Such an expansion allows for a faster convergence in numerical computations and for an angular analysis
of the BSA.

\begin{table}[t!]
\begin{center}
\begin{tabular}{c|c|c|c||c|c|c}
\hline \hline
      & $M_{P}$ &$M^\mathrm{exp}_P$   &  $\epsilon_{M_P}$ [\%]    &   $f_P $  &  $f^\mathrm{exp/lQCD}_P $  &  $\epsilon_{f_P} $ [\%]  \\ [0.5mm]  
           \hline
 $\pi (u \bar d)$&0.140&0.138&1.45 &   $0.094^{+0.001}_{-0.001}$ & 0.092(1)  & 2.17  \\ \hline
 $K (u \bar s)$  &0.494&0.494&0& $0.110^{+0.001}_{-0.001}$  &0.110(2)  &  0 \\\hline 
 $D (c \bar d) $  & $1.867^{+0.008}_{-0.004} $ &1.864 &0.11 &  $0.144^{+0.001}_{-0.001}$  &0.150 (0.5) & 4.00   \\\hline 
 $D_s(c \bar s) $ &  $2.015^{+0.021}_{-0.018}$ &1.968 & 2.39 & $0.179^{+0.004}_{-0.003}$  & 0.177(0.4) &1.13  \\\hline
 $\eta_c(c \bar c) $  & $3.012^{+0.003}_{-0.039}$  &2.984&0.94 &  $0.270^{+0.002}_{-0.005}$ &0.279(17) & 3.23 \\\hline
 $\eta_b(b \bar b) $  & $9.392^{+0.005}_{-0.004}$  &9.398 & 0.06 & $0.491^{+0.009}_{-0.009}$ &  0.472(4) & 4.03   \\\hline
 $B(u\bar b)$  &    $5.277^{+0.008}_{-0.005} $  & 5.279 & 0.04 & $0.132^{+0.004}_{-0.002}$  &0.134(1) &4.35   \\\hline
 $B_s(s\bar b)$  &  $5.383^{+0.037}_{-0.039}$   &5.367 &  0.30 &  $0.128^{+0.002}_{-0.003}$   &0.162(1) & 20.5 \\\hline
 $B_c(c\bar b)$ & $6.282^{+0.020}_{-0.024}$   & 6.274 & 0.13 &  $0.280^{+0.005}_{-0.002}$ & 0.302(2) & 10.17 \\
\hline \hline
\end{tabular}
\end{center}
\caption{Masses and decay constants [in GeV] of pseudoscalar mesons. All experimental masses and the pion and kaon weak decay constants are averaged values by the
Particle Data Group~\cite{ParticleDataGroup:2020ssz}. The leptonic decay constants of the $D_d$, $D_s$, $B_u$ and $B_s$ mesons are FLAG 2019 averages~\cite{Aoki:2019cca} 
and those of the $B_c$, $\eta_c$ and $\eta_b$ mesons  are from Ref.~\cite{McNeile:2012qf}. The relative deviations from experimental values, $v^\textrm{exp.}$, are  given by 
$\epsilon_v  =100\% \, | v^\textrm{exp.}  - v^\textrm{th.} | / v^\textrm{exp.}$. }     
\label{psproperties} 
\end{table}

We normalize the BSA with the Nakanishi condition~\cite{Nakanishi:1965zza} which involves the eigenvalue trajectory $\lambda (P^2)$ of the BSE solutions :
\begin{eqnarray}
  \left(\frac{\partial \ln (\lambda)}{\partial P^{2}}\right)^{-1}  \! =  \   \int \frac{d^4k}{(2\pi)^4}  \  \operatorname{tr_{CD}} \left [\bar{\Gamma}^{fg}_M (k ;-P) \,
        S_f  (k_{\eta } ) \, \Gamma^{fg}_M  (k ; P) S_g (k_{\bar \eta }) \right ] \ .
\label{nakanishinorm}                      
\end{eqnarray} 
The normalization at the mass pole, $P^2 = -M_P^2$, is required for the BSA  in the calculation of decay constants and other form factors or transition amplitudes. The weak decay constant of a 
pseudoscalar meson is defined by,
\begin{equation}
   f_P  P_\mu \ = \ \langle 0 | \bar q_g \gamma_5 \gamma_\mu q_f |  P  (P ) \rangle  \ ,
\end{equation}
and can be expressed by the integral:
\begin{equation}
\label{pseudodecay} 
  f_P  P_\mu  \ = \ \frac{\mathcal{Z}_2 N_c}{\sqrt{2} } \int^\Lambda\!  \frac{d^4k}{(2\pi)^4} \, \operatorname{Tr_D} \left [ i \gamma_5\gamma_\mu S_f (k_\eta)\, \Gamma_P^{fg} (k, P)\, S_g (k_{\bar \eta}) \right ] \, .
\end{equation}
As already noted,  the quark momenta, $k_\eta $ and $k_{\bar \eta} $,  define momentum-fraction parameters $\eta+\bar \eta =1$. Neither the decay constant nor any other physical observables can depend 
on them owing  to Poincar\'e covariance. The leptonic decay constant of a vector meson is defined by the amplitude,
\begin{equation}
   f_V  M_V  \bm{\epsilon}^\lambda_\mu   \ = \  \langle 0 | \bar q_g \gamma_\mu q_f |  V  (P,\lambda) \rangle  \ ,
\end{equation}
where $\bm{\epsilon}^\lambda_\mu (P)$ is the polarization vector of the transverse vector meson which satisfies  $\bm{\epsilon}^\lambda \cdot P = 0$ and is
normalized as ${\bm{\epsilon}^{\lambda}}^* \!\cdot \bm{\epsilon}^\lambda = 3$. This can again be expressed by a loop integral:
\begin{equation}
 f_V  M_V  \ =  \ \frac{\mathcal{Z}_2 N_c}{3\sqrt{2}}  \int^{\Lambda}\! \frac{d^4k}{(2\pi)^4} \,  \operatorname{Tr_D} \left [  \gamma_\mu  S_f (k_\eta)\,\Gamma_{V\mu}^{fg} (k, P) \, S_g  (k_{\bar \eta} ) \right ] \ .
\label{vectordecay}
\end{equation}

\begin{table}[t!]
\begin{center}
\begin{tabular}{c|c|c|c||c|c|c}
\hline \hline
& $M_V$ &$M^\mathrm{exp}_V$   &  $\epsilon_{M_V}$ [\%]    &   $f_V $  &  $f^\mathrm{exp/lQCD}_V $  &  $\epsilon_{f_V}$ [\%]  \\ [0.5mm]  
           \hline
$\rho (u\bar u)$&0.730&0.775&5.81 &0.145& 0.153(1) & 5.23\\
$\phi(s\bar s)$  &1.070&1.019&5.20&0.187&0.168(1)  &11.31 \\
$K^*(u\bar s)$  &0.942&0.896&5.13&0.177&0.159(1) &11.32 \\
$D^*(c \bar d) $&2.021&2.009&0.60&  0.165 &0.158(6) &4.43   \\
$D^*_s(c \bar s) $&2.169  &2.112 & 2.70& 0.205 & 0.190(5) &7.90  \\
 $J/\psi(c \bar c) $&3.124 &3.097 & 0.87&0.277 &0.294(5) &5.78 \\
 $\Upsilon(b \bar b) $&9.411 &9.460 & 0.52 &0.594&0.505(4) &17.62  \\
\hline \hline
\end{tabular}
\end{center}
\caption{ Masses and decay constants [in GeV] of ground-state vector mesons (preliminary without error estimate). The experimental masses 
are values listed by the Particle Data Group~\cite{ParticleDataGroup:2020ssz} and the leptonic decay constants for the $\rho$, $K^*$, $\phi$, $J/\psi$ and  $\Upsilon$ mesons are 
extracted from their experimental decay width~\cite{ParticleDataGroup:2020ssz} via $f_{V}^{2}=\frac{3 m_{V}}{4 \pi \alpha^{2} Q^{2}}\,  \Gamma_{V \rightarrow e^{+} e^{-}}$. 
The reference weak decay constants of the $D^*$ and $D^*_s$ mesons are those of the ETM collaboration~\cite{Lubicz:2017asp}. Relative deviations as in Table~\ref{psproperties}.}
\label{vecproperties}
\end{table}

Our results for the pseudoscalar and vector meson masses and leptonic decay constants are tabulated in Tables~\ref{psproperties} and \ref{vecproperties}. In the pseudoscalar
channel the experimental masses are reproduced within 1\%, while our calculated decay constants compare very well with experimental or theoretical reference values. We only note some
discrepancy in the case of  $f_{B_s}$ and $f_{B_c}$. We point out that the pion and kaon set the scale for the light- and strange-quark masses, respectively, and serve to fix 
the parameter combinations, $\omega_u D_u$ and $\omega_s D_s$. Therefore, no error estimate is give for these mesons, whereas the theoretical errors for the remaining meson masses 
are due to a certain insensitivity of the computed pion and kaon masses with respect to $\omega_u$ and $\omega_s$. The origins of theoretical errors and the choice of current-quark
masses at the renormalization scale $\mu$ in the DSE~\eqref{QuarkDSE} are discussed in detail in Ref.~\cite{Serna:2020txe}. The experimental $\rho$, $K^*$ and $\phi$ masses are also 
reasonably well reproduced and we observe very good agreement for the charmed mesons and heavy quarkonia. 

Finally, let us place emphasis on the fact that our results for the heavy-light  mesons hinge on our choice of a distinct flavor dependence in the interaction function~\eqref{BSEflavor}. Indeed, 
the strong asymmetry in the momentum distribution of the heavy and light quarks within these mesons requires a distinct treatment of their interaction with a gluon. Moreover, this ansatz also
facilitates the calculation of the quark propagators in the complex momentum plane in Euclidean space~\cite{Serna:2020txe}, where previous studies~\cite{Rojas:2014aka} were plagued
with cuts and singularities in the deep time-like domain.


\section{Conclusive remarks}

We have surveyed a selection of recent developments and outstanding issues over the past decade that have touched upon hadronic aspects of flavor physics. The common thread
of the different topics we addressed, be it effective Lagrangians and their effective couplings, flavor symmetry and its breaking patterns or the properties of flavored mesons and quarkonia, 
is nonperturbative QCD. Indeed, in computing couplings and form factors we necessarily deal with flavored antiquark-quark bound states. The nonperturbative aspects of bound states  
reveal themselves at different mass scales, where they must be considered individually as well as in conjunction with the other scales of a given physical problem. 

We also explored two symmetries, none of which is apparent in the QCD Lagrangian. Heavy-quark spin and flavor symmetries emerge only when the heavy sector of QCD is expanded in 
terms of inverse powers of the heavy-quark mass. We saw that  beyond-leading orders can already significantly contribute  to the breaking of these symmetries. Indeed, nonperturbative 
calculations of effective Lagrangian's couplings in a symmetry-preserving truncation of the DSE and BSE reveal that SU(3)$_F$ flavor symmetry is breached at the order of 20\%. 
While this may be acceptable for calculations with effective hadron degrees of freedom in the strange sector, the same cannot be said about SU(4)$_F$ as similar calculations demonstrate 
an order of magnitude larger symmetry-breaking effect. Along with the observation that the  charm quark is not a heavy enough quark, and therefore cannot be employed in a reliable expansion 
of HQET, we conclude that effective Lagrangians based on charmed meson degrees of freedom must be used judiciously. In particular, flavor-symmetry breaking in the effective couplings 
should be accounted for. 

At last, we discussed how reliable calculations of heavy-light form factors, couplings and decay constants require the correct description of the light quark's propagator as solutions of the DSE 
or gap equation, while the meson's wave functions are obtained from Poincar\'e-invariant solutions of the corresponding BSE. The couplings we discussed in here were mostly obtained with 
simplified models of these wave function. Their  calculation with the heavy-light BSAs presented in Section~\ref{sec4} is in progress and results will be available shortly.

\acknowledgments

While the ongoing pandemic unfortunately didn't allow for an on-site meeting in Mexico City, we thank the organizers for the opportunity of  a ``virtual presentation'' and a well-organized
on-line conference. This work is supported by the Brazilian agencies FAPESP, grant no.~2018/20218-4, and CNPq, grant no.~428003/2018-4, and is part of the project ``INCT-F\'isica Nuclear 
e Aplica\c{c}\~oes'',  no. 464898/2014-5. F.E.S. is a CAPES-PNPD postdoctoral fellow, contract no.~88882.314890/2013-01.



\end{document}